\documentclass[twocolumn]{aastex62}

\usepackage{graphicx}
\usepackage{xspace}
\usepackage{color}
\usepackage{hyperref}
\usepackage{listings}
\usepackage[]{algorithm2e}
\usepackage{amsmath}
\usepackage{amssymb}
\usepackage{mathrsfs}
\usepackage{mathtools}
\usepackage{dashrule}
\usepackage{lineno}

\makeatletter
\long\def\frontmatter@title@above{
\ifrnaas
\vspace*{-\headsep}\vspace*{\headheight}
\footnotesize
\noindent{ }\\[2pt]
{\footnotesize Typeset using \LaTeX\ {\bf RNAAS} style in AASTeX62}
\par\vspace*{-\baselineskip}\vspace*{0.625in}
\else
\vbox to 0pt{\vskip-48pt\normalsize\rptloopnum=0\loop
\global\advance\rptloopnum by 1
\expandafter\ifx\csname report\the\rptloopnum\endcsname\relax
\else
\csname report\the\rptloopnum\endcsname
\vskip1pt
\repeat
\vss
}
\ifmodern
\vglue-18pt
{\footnotesize
\raggedright
{ }\\[2pt]
{\footnotesize
Typeset using \LaTeX\ {\bf modern} style in AASTeX62}
\vskip48pt
}
\else
\ifpreprint
\vspace*{-\headsep}\vspace*{\headheight}
\footnotesize
\noindent{ }\\[2pt]
{\footnotesize Typeset using \LaTeX\ {\bf preprint} style in AASTeX62}
\par\vspace*{-\baselineskip}\vspace*{0.625in}
\else
\ifpreprinttwo
\vspace*{-\headsep}\vspace*{\headheight}
\footnotesize
{\footnotesize\textsc{\@journalinfo}}\par
{\footnotesize Typeset using \LaTeX\ {\bf preprint2} style in AASTeX62}
\par\vspace*{-\baselineskip}\vspace*{0.625in}
\else
\iftwocolstyle
\vspace*{-\headsep}\vspace*{\headheight}
\footnotesize
{\footnotesize\textsc{\@journalinfo}}\par
{\footnotesize Typeset using \LaTeX\ {\bf twocolumn} style in AASTeX62}
\par\vspace*{-\baselineskip}\vspace*{0.625in}
\else
\ifmanu
\vspace*{-\headsep}\vspace*{\headheight}
\footnotesize
\noindent\textsc{\@journalinfo}\\[-8pt]
{\footnotesize Typeset using \LaTeX\ {\bf manuscript} style in
AASTeX62}
\par\vspace*{-\baselineskip}\vspace*{0.625in}
\else
\vspace*{-\headsep}\vspace*{\headheight}
\footnotesize
{\footnotesize\textsc{\@journalinfo}}\par
{\footnotesize Typeset using \LaTeX\ default style in AASTeX62}
\par\vspace*{-\baselineskip}\vspace*{0.625in}
\fi\fi\fi\fi\fi
\fi 
}%
\def\journalinfo#1{\gdef\@journalinfo{#1}}
\journalinfo{ }
\def\titleblock@produce{%
 \begingroup
 \ltx@footnote@pop
 \def\@mpfn{mpfootnote}%
 \def\thempfn{\thempfootnote}%
 \c@mpfootnote\z@
 \let\@makefnmark\frontmatter@makefnmark
  \frontmatter@setup
  \thispagestyle{titlepage}\label{FirstPage}%
\ifmodern\leftskip=0pt\rightskip\leftskip\fi
  \frontmatter@title@produce
  \groupauthors@sw{%
\frontmatter@author@produce@group
  }{%
   \frontmatter@author@produce@script
  }%
  \frontmatter@RRAPformat{ }%
\expandafter\ifx\csname @submitted\endcsname\relax\else
\vskip6pt
\expandafter\produce@RRAP\expandafter{\centerline{\@submitted\hbox
to 20pt{\hfill}}\vskip12pt}%
\fi
  \frontmatter@abstract@produce
  \@ifx@empty\@pacs{}{%
   \@pacs@produce\@pacs
  }%
  \@ifx@empty\@keywords{}{%
   \@keywords@produce\@keywords
  }%
  \par
  \frontmatter@finalspace
\endgroup%
}%

\makeatother

\graphicspath{{./}{figures/}}

\received{\today}
\revised{TBD}
\accepted{TBD}

\shorttitle{Gaussbock: Fast cosmological parameter estimation}

\shortauthors{Moews \& Zuntz}

\begin{document}

\title{Gaussbock: Fast parallel-iterative cosmological parameter estimation with Bayesian nonparametrics}

\correspondingauthor{Ben Moews}
\email{bmoews@roe.ac.uk}

\author[0000-0003-0897-040X]{Ben Moews}
\affil{Institute for Astronomy, University of Edinburgh, Royal Observatory, Edinburgh EH9 3HJ, UK}

\author[0000-0001-9789-9646]{Joe Zuntz}
\affil{Institute for Astronomy, University of Edinburgh, Royal Observatory, Edinburgh EH9 3HJ, UK}

\begin{abstract}
We present and apply Gaussbock, a new embarrassingly parallel iterative algorithm for cosmological parameter estimation designed for an era of cheap parallel computing resources. Gaussbock uses Bayesian nonparametrics and truncated importance sampling to accurately draw samples from posterior distributions with an orders-of-magnitude speed-up in wall time over alternative methods. Contemporary problems in this area often suffer from both increased computational costs due to high-dimensional parameter spaces and consequent excessive time requirements, as well as the need for fine tuning of proposal distributions or sampling parameters. Gaussbock is designed specifically with these issues in mind. We explore and validate the performance and convergence of the algorithm on a fast approximation to the Dark Energy Survey Year 1 (DES Y1) posterior, finding reasonable scaling behavior with the number of parameters. We then test on the full DES Y1 posterior using large-scale supercomputing facilities, and recover reasonable agreement with previous chains, although the algorithm can underestimate the tails of poorly-constrained parameters. Additionally, we discuss and demonstrate how Gaussbock recovers complex posterior shapes very well at lower dimensions, but faces challenges to perform well on such distributions in higher dimensions. In addition, we provide the community with a user-friendly software tool for accelerated cosmological parameter estimation based on the methodology described in this paper.
\end{abstract}


\raggedbottom

\section{Introduction} \label{sec:intro}
Bayesian methods are now a standard approach to data analysis and inference in astrophysics. In this approach, probabilities are regarded as a means of quantifying information, and in particular the information contained in an experimental dataset about a specific model. This is encoded in the \textit{posterior}, which combines \textit{prior}, or external, information with the \textit{likelihood} from the current data. For textbooks providing an introduction and overview of Bayesian methods, we refer interested readers to \citet{Bernardo1994}, \citet{MacKay2003} , and \citet{Gelman2013}, as well as \citet{Murphy2012} and \citet{Hobson2009} for an overview centered on machine learning and cosmology, respectively. In most realistic cases, the analytic or direct numerical evaluation of posterior probability distributions is impossible or infeasible, especially in cases that feature many parameters, due to the large volume of high-dimensional spaces. The wide-spread use of Bayesian methods has largely been driven by the availability of \textit{sampling} algorithms, which can generate samples from a posterior distribution without exploring the full space. These samples can then be used to generate summary statistics like means and limits on individual parameters, or correlations between them. For a shorter overview of the application of Bayesian inference in cosmology, see \citet{Trotta2008}.

Within the cosmology literature, \citet{Christensen2001} proposed initial arguments for the use of Bayesian methods for the purpose of cosmological parameter estimation. They argued for \textit{Markov chain Monte Carlo} (MCMC) approaches due to their superiority in terms of sampling from, and converging to, the true posterior distribution in the limit of an infinite sample size. The application of MCMC approaches in these early efforts were centered on the \textit{Metropolis-Hastings algorithm}, which was named after work done by \citet{Metropolis1953} and, for the more general case, \citet{Hastings1970}. The distinguishing feature of this method is the acceptance of new points in the Markov chain if the likelihood ratio of the proposed point and the last point is larger than one, and the probabilistic acceptance of points with a lower ratio if the latter is larger than a random number $n \in [0, 1]$. This acceptance of less likely points dependent on the likelihoods leads to the sampling from the posterior distribution and, notably, does not require marginalization via the evidence. \citet{Knox2001} then followed the proposal of \citet{Christensen2001} to constrain the age of the universe to $t_0 = 14.0 \pm 0.5$ Gyr. Earlier work includes \citet{Saha1994}, who made use of the Metropolis-Hastings algorithm for for galaxy kinematics, and \citet{Christensen1998}, who employed the related Gibbs sampler for gravitational wave analysis \citep{Geman1984}. For more in-depth information covering the wide array of contributions from both the astrophysical and statistical literature, we recommend \citet{Trotta2008} as a more complete overview of the development of Bayesian inference in cosmology in particular, and \citet{Robert2011} and \citet{Brooks2011} for a history of MCMC methods and their development in general.

Up to, and into, the new millennium, the Metropolis-Hastings algorithm remains the standard approach to cosmological parameter estimation, which was further supported by the development of a dedicated implementation in \texttt{CosmoMC} \citep{Lewis2002}. A variety of algorithms and codes are, however, available for different types of problems. The optimal choice depends on multiple factors, including the dimensionality of the problem, meaning the number of parameters to estimate, the evaluation speed, the need for Bayesian evidences, the availability of analytic derivatives, the ability to sample from marginal distributions, and the possibility and degree of using parallelization.

In more recent years, new MCMC sampling techniques were proposed and subsequently applied to cosmological parameter estimation. Examples include \textit{Population Monte Carlo} (PMC) techniques introduced by \citet{Cappe2004} and \citet{Wraith2009}, and used by \citet{Kilbinger2010} to develop \texttt{CosmoPMC}; \textit{affine-invariant MCMC ensembles} by \citet{Goodman2010}, which led to the publication of \texttt{emcee} by \citet{Foreman-Mackey2013} and \texttt{CosmoHammer} by \citet{Akeret2013}; and renewed interest in \textit{Approximate Bayesian Computation} (ABC) for likelihood-free inference based on simulations to introduce \texttt{CosmoABC} and \texttt{abcpmc} \citep{Ishida2015, Akeret2015}. \textit{Density estimation likelihood-free inference} (DELFI) is a recently developed technique that trains a flexible density estimator to approximate the target posterior, circumventing the large number of simulations that traditional ABC approaches can require \citep{Bonassi2011, Fan2013, Papamakarios2016}. Using the JLA sample of 740 type SN Ia supernovae as described in \citet{Betoule2014}, \citet{Alsing2018} subsequently deploy this method to estimate cosmological parameters. Their approach, however, makes a few simplifying assumptions, for example normally distributed priors and likelihoods. Other advanced methods, like the \textit{Hamiltonian Monte Carlo} approach developed by \citet{Duane1987}, have also been applied, for example by \citet{Hajian2007}. These developments are driven by the computationally costly likelihood calculations involved in most MCMC algorithms, trying to alleviate this issue with a certain degree of parallelization due to the increased availability of cheap computing resources, faster convergence or, in the case of ABC, the circumvention of direct likelihood computations altogether.

As such methods either fail to reduce the runtime enough for modern problems or have their own pitfalls, for example through an increased risk of introducing biases, the quest for highly parallelized and fast alternatives for cosmological parameter estimation continues. This need is further exacerbated by upcoming missions like LSST and Euclid requiring high-dimensional posterior approximations with a large number of required nuisance parameters predicted to vastly exceed previous missions \citep{Amendola2018}.

It should be noted that the statistical literature on sampling methods is rich and vast, and a complete review of both their history and all current developments would exceed the scope of this paper. The methods covered in more detail here are those likely to be more familiar to the astrophysical community, due to being wide-spread or featuring field-specific implementations. While we aim to cover relevant comparisons, this should, of course, not be misunderstood as a judgment about these methods being superior in the wider context of all statistical developments, but to place this work in the context of astrostatistics.

\textit{Nested sampling} is a Bayesian take on numerical Lebesque integration for model selection introduced by \citet{Skilling2006}. While targeting the calculation of Bayesian evidence, posterior samples are generated as a by-product, and the algorithm was quickly shown to require considerably fewer posterior evaluations \citep{Mukherjee2006}. Due to denser and sparser sampling from high-posterior and low-posterior regions, respectively, nested sampling provides increased efficiency when compared to previous MCMC methods. This has lead to extensions and implementations for applications in cosmology, notably \texttt{CosmoNest} by \citet{Liddle2006}, \texttt{MultiNest} as described in \citet{Hobson2008} and \citet{Feroz2009}, and \texttt{PolyChord} \citep{Handley2015}. In cosmology, such implementations have been used in areas as diverse as cosmic ray propagation models, cosmoparticle physics, and gravitational wave astronomy \citep{Trotta2011, DelPozzo2012, Verde2013, DelPozzo2017, Wang2018}. A comparison between nested sampling and state-of-the-art MCMC methods can be found in \citet{Allison2014}, while an investigation of statistical uncertainties in nested sampling is provided by \citet{Keeton2011}. Nested sampling has also been adopted by other fields of research, including GPU-accelerated implementations, for example in systems biology \citep{Aitken2013, Stumpf2014}.

The statistical literature, however, points out various issues of nested sampling methods that have prevented wide-spread adoption in statistics. Among these are the assumption that perfect and independent samples from a constrained version of the prior are drawn in each iteration, the underestimation of sampling errors due to the simulated-weights method it employs, and an asymptotic approximation variance that scales linearly with the dimensionality of a given parameter space \citep{Chopin2010, Higson2018}.

In this paper, we use example likelihoods from the Dark Energy Survey (DES) collaboration's analysis of lensing and clustering data, as presented in \citet{Abbott2018a}. These calculations make use of the \texttt{CosmoSIS} and \texttt{CosmoLike} pipelines, which contain implementations of both \texttt{Multinest} and \texttt{emcee} \citep{Zuntz2015, Krause2017}.

For a comparison of approaches designed for the acceleration of MCMC methods in particular, including additional parallelization methods, see \citet{Robert2018}, who cover methods targeting both the exploration stage of the algorithms and the exploitation level. The second approach includes Rao-Blackwellization and scalability, with the latter encompassing parallelization under this nomenclature. Other examples of methods trying to optimize the performance of established algorithms include the \textit{no-U-turn sampler} (NUTS) by \citet{Hoffmann2014}, which alleviates the need of the previously mentioned HMC algorithm for tuning by computing the trajectory length via recursively built candidate proposals, as well as work by \citet{Neiswanger2014} on asymptotically exact and embarrassingly parallel MCMC sampling. The latter solves the slowing-down of parallel MCMC methods by reducing the amount of required communication in a divide-and-conquer tactic that splits up the dataset and which the authors justify with prohibitively long runtimes of many serial methods. Interestingly, our method allows for the use of any sampling method to create the initial sample, meaning that such optimized divide-and-conquer methods can be easily incorporated into our approach. The need for sped-up posterior estimation approaches is further elaborated on by \citet{Bardenet2014} and \citet{Wilkinson2005}, with the latter pointing out the need for parallelized method due to: ``[...] weeks of CPU time on powerful computers'' for serial MCMC methods on high-dimensional problems of interest. The need for parallelization approaches stems mostly from cases in which parts of the computations are very expensive, but which can be transformed into an, ideally, embarrassingly parallel problem that allows the respective steps to take full advantage of a greater number of cores, thus cutting otherwise infeasible runtimes to a fraction. For a more general overview of both the history and more recent developments in the field of Monte Carlo methods, such as multi-stage Gibbs samplers, see \citet{Robert2004}.

In this paper, we propose a parallel-iterative algorithm to address these issues, making use of recent advances in the fields of statistics and machine learning. Our method starts with a preliminary approximation of the target distribution, either through a built-in affine-invariant MCMC ensemble or a user-provided initial sample guess. We then fit a non-parametric model to the sample and employ a variation of sampling-importance-resampling to iteratively move the samples toward the true distribution, repeating these steps until the process converges. In doing so, we also offer a user-friendly Python package to both the cosmology and the wider astronomy community, as well as a general parameter estimation tool for other disciplines dealing with the same issues. We test our implementation on the DES Year 1 (Y1) posterior, and on a fast approximation to the latter for extended tests.

The remainder of this paper is structured as follows: We cover the relevant methodology, which includes an overview of variational inference for Bayesian mixture models and truncated importance sampling, as well as the mathematical architecture of the proposed approach, in Section~\ref{sec:methodology}. In Section~\ref{sec:implementation}, we introduce an open-source implementation based on our method, explain computational considerations and parallelization, and provide a quickstart tutorial. Experiments for both toy examples and an approximation of the DES Y1 likelihood are covered in Section~\ref{sec:experiments}, together with cosmological parameter estimation runs on supercomputing facilities for the full DES Y1 likelihood. We present and discuss the results of these experiments in Section~\ref{sec:results}, and summarize our findings in the conclusion in Section~\ref{sec:conclusion}.

\newpage

\section{Mathematical background} \label{sec:methodology}

While Bayesian inference has earned its place as a powerful instrument for cosmological research, complex problems often suffer from the need to approximate probability densities that are difficult or outright infeasible to compute. Since Bayesian methods rely on the posterior density, approximations are a necessary evil. In the algorithm presented in this paper, we fit a mixture model to sample from the posterior using \emph{variational inference} methods, while avoiding fixing the number of mixture components by using a \emph{Dirichlet process}. We iteratively improve these samples using \emph{truncated importance sampling} until a convergence criterion is fulfilled.

In this section, we introduce variational inference in Section~\ref{sec:variational}, followed by Dirichlet processes and the stick-breaking procedure in Section~\ref{sec:dirichlet}. After an overview of sampling-importance-resampling and truncated importance sampling in Section~\ref{sec:truncated}, we introduce a novel method for parallel-iterative parameter estimation in Section~\ref{sec:estimation}.

\subsection{Variational inference for mixture models} \label{sec:variational}

Variational Bayesian methods were originally developed and explored in the context of artificial neural networks, and gained initial interest from research on inference in graphical models \citep{Peterson1987, Peterson1989, Jordan1999}. The use of variational Bayesian methods for inference is commonly known as \textit{variational inference} (VI) and provides a faster and more scalable alternative to Markov chain Monte Carlo (MCMC) methods in many contexts; the main difference between them is that VI treats parameter estimation not as a sampling problem, but instead as an optimization problem. From a research point of view, these methods also garnered the interest of the statistics community because they are currently not as well understood as MCMC methods \citep{Blei2017}. 

The \textit{Kullback-Leibler divergence} $D_{\text{KL}}$ is a central concept in VI and defines by how much a distribution diverges from another, or how similar it is. For a reference distribution $p(\textbf{x})$ and a proposal distribution $q(\textbf{x})$, the $D_{\text{KL}}$ can be expressed as
\begin{equation}
D_{\text{KL}}(p(\textbf{x}) || q(\textbf{x})) = \int^{\infty}_{-\infty} p(\textbf{x}) \text{log} \frac{p(\textbf{x})}{q(\textbf{x})} \ d \textbf{x}.
\end{equation}
The fact that the $D_{\text{KL}}$ is an asymmetric difference measure means that $D_{\text{KL}}(p(\textbf{x}) || q(\textbf{x})) \neq D_{\text{KL}}(q(\textbf{x}) || p(\textbf{x}))$, which is due to its calculation as a directional loss of information. 

In VI, the $D_{\text{KL}}$ is used to find a best-fitting distribution to a set of samples. Let $\mathscr{Q}$ be a selected family of distributions, $\textbf{x}$ and $\textbf{z}$ observations and parameters, respectively, and $p(\textbf{z})$ a prior density that can be related to observations via the likelihood $p(\textbf{x} | \textbf{z})$ to calculate the posterior $p(\textbf{z} | \textbf{x})$. The family member $\hat{q}(\textbf{z})$ that best matches the posterior can be found in the framework of an optimization problem, finding with some specified tolerance the value of
\begin{align}
\begin{split}
\hat{q}(\textbf{z}) &= \underset{q(\textbf{z}) \in \mathscr{Q}}{\operatorname{argmin}} \ D_{\text{KL}} \left( q(\textbf{z}) || p(\textbf{z} | \textbf{x}) \right).
\end{split}
\end{align}
Calculating this quantity directly is often infeasible, since it is equivalent to measuring the Bayesian evidence. Instead, VI methods (equivalently) maximize an alternative quantity, the \textit{evidence lower bound} (ELBO),
\begin{align}
\begin{split}
\text{ELBO} (q) &= \mathbb{E} [\text{log } p(\textbf{z}, \textbf{x})] - \mathbb{E} [\text{log } q(\textbf{z})]\\
 &= \mathbb{E} [\text{log } p(\textbf{x} | \textbf{z})] - D_{\text{KL}} \left( q(\textbf{z}) || p(\textbf{z}) \right),
\end{split}
\end{align}
which is numerically easier to calculate than the $ D_{\text{KL}}$. The ELBO also delivers a lower bound for the evidence, which is the reason for the utility of VI for model selection, as covered in \citet{Blei2017}. A more extensive introduction to VI for the interested reader can be found in \citet{Murphy2012}.

\subsection{Dirichlet processes and stick-breaking} \label{sec:dirichlet}

Instead of the more traditional approach of fixing the number of components in the mixture model that we use to model the posterior, we determine the component number from the sample itself at each iteration. This approach employs a \textit{Dirichlet Process} (DP) as a prior on the number of parameters, which enables the use of a suitable number of components during each step, meaning that changes between iterations are not forced to use the same components.

Developed by \citet{Ferguson1973}, DPs are distributions of distributions, featuring a base distribution $G_0$ and a scaling parameter $\alpha \in \mathbb{R}_+$, and with realizations denoted as $G \sim \text{DP}(\alpha, G_0)$. This area has important applications as the prior in infinite mixture models, and gained new traction in both statistics and machine learning in recent years \citep{Gershman2012}. The DP mixture model presented originally by \citet{Antoniak1974} takes $\theta_i$ as the distribution parameter of observation $i$ and uses the discrete nature of the base distribution $G_0$ to view the DP mixture as an infinite mixture model. For samples \textbf{s} from such a DP mixture, with sample size $N$, the predictive density with $\mathbf{s} = s_1, ... , s_N$ is
\begin{equation}
p(s|\mathbf{s}, \alpha, G_0) = \int p(s|\theta) p(\theta|\mathbf{s}, \alpha, G_0) \ d \theta.
\end{equation}
As the computation of that density is, again, infeasible, \citet{Blei2006} introduce the use of VI for DP mixtures. Bayesian takes on mixture models employ a prior over the mixing distribution as well as over the cluster parameters, with the former commonly being a Dirichlet and the latter being a Gaussian distribution in our case. Given the discrete nature of random measures drawn from a DP, a mixutre of the latter can be viewed as a mixture model with an unbounded number of components \citep{Blei2006}.

The Bayesian nonparametrics approach employs the \textit{stick-breaking process} by \citet{Sethuraman1994}, which exploits the discrete nature of DPs to calculate the probability mass function, and can be used for Bayesian Gaussian mixtures with an undetermined number of Gaussians. The name is based on the analogy of breaking a stick of unit length into infinite segments by consecutively breaking off $\beta_1$, $\beta_2$, etc. from the stick until the remainder is truncated to recover a finite-dimensional representation. The truncated variational distribution is then used to approximate the posterior of an infinite DP mixture. As a mathematical description of the subsequent application of VI to DPs with stick-breaking would go beyond the scope of this overview, we refer the reader to \citet{Blei2006}. A less concise introduction to DPs and Bayesian nonparametrics in general, as well as its applications, is provided in \citet{Hjort2010}.

As the posterior distribution, given a DP mixture prior, cannot be directly calculated, VI offers a deterministic approach to approximate them. In this paper, we employ the mean-field family within VI to optimize the $D_{\mathrm{KL}}$, using this approach to approximate the joint posterior for parameters of an infinite Gaussian mixture, made finite to a maximum number of components through stick-breaking.

\subsection{Importance Sampling} \label{sec:importance}

\textit{Importance sampling} was described early by \citet{Kahn1953} in the context of sample size reduction in Monte Carlo methods and continues to inspire a wide array of extensions. This includes physics-specific techniques like umbrella sampling for difficult energy landscapes by \citet{Torrie1977} and, more recently, methods to alleviate issues with poorly approximated proposal distributions \citep{Ionides2008}. It is also a staple in cosmological parameter estimation, for example in \citet{Wraith2009} and \citet{Kilbinger2010}. Generally, the basic method is a way to estimate distribution properties if only samples from a different, often approximated, distribution are given. Let $p(\textbf{z})$ be the target distribution, $q(\textbf{z})$ an approximate (or proposed) distribution, and $f(\textbf{z})$ some function. The expectation of $f(\textbf{z})$ can then be computed as
\begin{align}
\begin{split}
\mathbb{E}[f] &= \int f(\textbf{z}) p(\textbf{z}) \ d\textbf{z}\\
 &= \int f(\textbf{z}) \frac{p(\textbf{z})}{q(\textbf{z})} q(\textbf{z}) \ d\textbf{z}\\
 &\simeq \ \frac{1}{N} \sum^N_{i = 1} \frac{p(\textbf{z}_i)}{q(\textbf{z}_i)} f(\textbf{z}_i),
\end{split}
\end{align}
with $N$ as the number of drawn samples. The ratios in this equation, given as
\begin{equation}
r_l \equiv \frac{p(\textbf{z}_l)}{q(\textbf{z}_l)},
\end{equation}
are called the \emph{importance weights} or \emph{importance ratios} and are central to the method.

\textit{Sampling-importance-resampling} (SIR) is a two-step approach in which the importance weights for a set of samples are calculated, after which an equally-sized subset of these samples is generated by drawing from them with probabilities per sample indicated by the normalized importance weights. For a more in-depth introduction to importance sampling and other related sampling methods, see \citet{Bishop2006}.

\subsection{Counteracting high-weight samples} \label{sec:truncated}

One issue with this approach is the possibility of overly dominant samples, meaning points with disproportionately high posterior values in comparison to the rest of a set of model samples. During the importance resampling step, this dominance leads to copies of these samples being overrepresented, resulting in sets that are too narrow in their densities. We address this issue with \textit{truncated} importance sampling, an extension of importance sampling that truncates weights of high-value samples based on the total number of drawn samples, with guarantees for finite variance and mean-square consistency under weak conditions \citep{Ionides2008}. For a set of $N_i$ samples, proposal distribution posteriors $q(\theta_i)$, actual posteriors $p(\theta_i)$ and a set truncation value $\alpha$ with justifications to be set at $\alpha = 2$, the weight $w_i$ of a single sample is updated according to
\begin{equation}
w_i = \text{min}\left( r_i, \bar{r} N_i^{\frac{1}{\alpha}} \right) \mathrm{ , with} \ r_i = \frac{p(\theta_i)}{q(\theta_i)},
\end{equation}
where $\bar{r}$ is the mean of all importance weights for the sample. With this extension applied to SIR, the weighted drawing of samples is limited by the truncation value. This change improves the behavior of importance sampling during the early part of the algorithm described below, when the estimated distribution $q$ is a poor approximation to the desired posterior $p$, and alleviates the issue of working with relatively small sample sizes for high-dimensional parameter spaces.

\section{The Gaussbock Algorithm} \label{sec:estimation}

Based on Bayesian nonparametrics and machine learning as described in Sections~\ref{sec:variational} and \ref{sec:dirichlet}, we introduce an algorithm that uses variational inference on an infinite Dirichlet process approximated via stick-breaking to fit variational Bayesian Gaussian mixture models (GMMs) in an iterative manner. This algorithm offers a highly adaptive and embarrassingly parallel way to approximate high-dimensional posteriors with computationally expensive likelihoods.

\begin{algorithm}[!htbp]
    \label{alg:gaussbock}
    \caption{Pseudo-code for \texttt{Gaussbock}.}
	\KwData{Initial posterior-space samples $\boldsymbol{\theta}_{\textup{start}}$,\\
	\hspace{30pt}number of required output samples $n$,\\
	\hspace{30pt}array of allowed ranges per parameter $\mathbf{r}$,\\
	\hspace{30pt}number of samples drawn per iteration $m$,\\
	\hspace{30pt}safety margin multiplier for sampling $c$,\\
	\hspace{30pt}maximum number of mixture components $g$,\\
	\hspace{30pt}dynamically shrinking fitting tolerance $d$,\\
	\hspace{30pt}value for importance weight truncation $\alpha$,\\
	\hspace{30pt}log-posterior function for $p(\boldsymbol{\theta} | \mathcal{D})$}
	\KwResult{Approximated posterior samples $\boldsymbol{\theta}_{\textup{final}}$}
	$\boldsymbol{\theta}_{\textup{new}} \leftarrow \boldsymbol{\theta}_{\textup{start}}$\;
	\For{$i \gets 1$ \textbf{to} $N$}{
		\textit{Calculate the (shrinking) model fitting tolerance}\\
		$d \leftarrow a_1 - i \cdot \Delta a \cdot (N - 1)^{-1} $\\
		\textit{Fit a variational Bayesian GMM to the samples}\\
		$\mathcal{M}_i \leftarrow \textup{VBGMM}(\boldsymbol{\theta}_{\textup{new}}, d, g)$\\
		\textit{Sample a set of parameters from the fitted model}\\
		$\boldsymbol{\theta}_i \leftarrow \boldsymbol{\theta} \sim \mathcal{M}_i \textup{ s.t. } \textup{length}(\boldsymbol{\theta}) = m \cdot c$\\
		\textit{Cut samples straying beyond the allowed ranges}\\
		$R = \textup{r}_1 \times \textup{r}_2 \times \ldots \times \textup{r}_{\dim(\boldsymbol{\theta}_i)}$\\ 
		$\boldsymbol{\theta}_i \leftarrow \boldsymbol{\theta}_i \cap R$\\
		\textit{Keep the required number of parameter samples}\\
		$\boldsymbol{\theta}_i \leftarrow \boldsymbol{\theta}_i^{(1:n)}$\\
		\textit{Parallel calculation of true log-posterior values}\\
		$\textbf{p} \leftarrow p(\boldsymbol{\theta}_i | \mathcal{D})$\\
		\textit{Compute importance probabilities in linear space}\\
		$\textbf{w}_i \leftarrow \textup{exp}(\textbf{p} - p(\boldsymbol{\theta}_i | \mathcal{M}))$\;
		\textit{Compute the truncated importance probabilities}\\
		$\textbf{w}_i \leftarrow \textup{min}(\textbf{w}_i, \bar{\textbf{w}}_i \cdot \textup{length}(\boldsymbol{\theta}_i^{\frac{1}{\alpha}}))$\\
		\textit{Renormalize the updated importance probabilities}\\
		$\textbf{w}_i \leftarrow \textbf{w}_i \times \left(\sum \textbf{w}_i\right)^{-1}$\;
		\textit{Weighted sampling from the parameter samples}\\
		$L \leftarrow \textup{length}(\boldsymbol{\theta}_i)$\\
		$\boldsymbol{\theta}_{\textup{new}} \leftarrow \textup{sample}(\boldsymbol{\theta}_i, \textbf{w}_i) \textup{ s.t. } \textup{length}(\boldsymbol{\theta}_{\textup{new}}) = L$\\
		\textit{Terminate if convergence criterion is reached}\\
		\If{$| \Delta \sigma_i^2 < t$}{
		\textup{\textbf{break}}\\
		}
	}
	\textit{Return the user-specified number of final samples}\\
	\Return $\boldsymbol{\theta}_{\textup{final}} \leftarrow \boldsymbol{\theta} \sim \mathcal{M}_i \textup{ s.t. } \textup{length}(\boldsymbol{\theta}) = n$\\
	\vspace{10pt}
\end{algorithm}

The idea behind our approach is to start from an initial sample guess, either from existing data or a short run of another sampler such as \texttt{emcee}. Based on the work on nonparametric VI by \citet{Gershman2012}, our algorithm uses a variational Bayesian GMM due to its ability to automatically determine the number of Gaussians required to produce a good fit by stick-breaking an infinite Dirichlet process mixture. For this reason, only the provision of a maximal number of Gaussians is required. The algorithm then determines means and variances for the optimal number of Gaussians given a sample and fitting tolerance. This is followed by drawing a new sample from the fitted mixture model, and a truncated SIR step to move the sample distribution further toward the true the posterior density. These steps are then repeated in an iterative manner until convergence, which is assessed from the change in the variance of importance weights at the end of each iteration: 
\begin{enumerate}
\item Fit a variational Bayesian GMM to the sample,
\item draw a new sample from the newly fitted model,
\item perform an SIR step for a weighted sample, and
\item check inter-iteration variances for convergence.
\end{enumerate}
We use a dynamically shrinking tolerance $d$ for the model-fitting step. Let $a$ be the tuple denoting the initial and final model-fitting tolerances, with $a_1 > a_2$, and let $N$ be the maximum number of iterations, then the tolerance $d_i$ for a given iteration $i \in \{1, 2, \dots, N\}$ is
\begin{equation}
d_i = a_1 - i \cdot \Delta a \cdot (N - 1)^{-1} \mathrm{, \ with \ } \Delta a = a_1 - a_2.
\end{equation}
This approach is related to the previously mentioned PMC algorithms initially introduced by \citet{Cappe2004}, and applied to cosmological inference in \citet{Kilbinger2010}. It differs, though, by the nonparametric nature of the model, which eliminates the bias present in the predetermined number of distributions in classical GMMs. It also adds the weight truncation to reduce the influence of overly dominant samples with high posterior values in relatively small samples. Our method bears motivational similarity, although considerable methodological differences, to \texttt{CosmoABC}, while not being subject to the potential pitfalls of forward-simulation inference in ABC \citep{Ishida2015}.

In Algorithm~\ref{alg:gaussbock}, we provide a more complete pseudo-code representation of the most relevant parts of the approach described in this paper, which we name \texttt{Gaussbock}. For this algorithm, we let $\mathbf{r}$ be the array of tuples representing the allowed ranges (min, max) per dimension, that is, per parameter. Furthermore, let $N$ be the maximum number of iterations, $m$ the number of samples to be drawn from each iteration's model, $n$ the number of samples returned after termination, $g$ the maximum number of Gaussians available for approximating the posterior distribution, and $c$ a safety margin parameter greater than one to draw additional GMM samples in case some fall outside the parameter bounds. Finally, let $\mathcal{D}$ be the empirical data used for calculating the true likelihood. The specifics of the variational Bayesian GMM (VBGMM) with reasonable default settings, like the prior of the covariance distribution and the parameter initialization for the VBGMM, are omitted in order to keep the pseudo-code concise.

As (mostly adaptive) defaults are used for the settings of \texttt{Gaussbock}, only the initial approximative sample set $\boldsymbol{\theta}_{\textup{start}}$, the number of iterations $n$, and the handle of a function to compute $p(\boldsymbol{\theta}_i | \mathcal{D})$ have to be provided with regard to the above pseudo-code. In addition, if no $\boldsymbol{\theta}_{\textup{start}}$ is provided, the implementation described in Section~\ref{sec:implementation} will automatically run an affine-invariant MCMC ensemble to procure that initial set of posterior-space samples. Since the determination of convergence is a common issue in MCMC methods, \texttt{Gaussbock} uses a convergence threshold $t$ that terminates the iterative fitting-resampling procedure if reached before the maximum number of iterations. For this purpose, we measure the difference in inter-iteration weight variances $\Delta \sigma_i^2$, which takes the form
\begin{equation}
\begin{split}
&\Delta \sigma_i^2 = | \bar{\sigma}_i^2 - \bar{\sigma}_{i - 1}^2 |,\\ &\mathrm{with \ } \bar{\sigma}_i^2 = \dim{(D)}^{-1} \sum\limits_{d = 1}^{\dim{(D)}} \sigma (\log (w_{i_d}))^2.
\end{split}
\end{equation}
Here, the average logarithmic importance weight variance is denoted as $\bar{\sigma}_i^2$, providing the arithmetic mean over the dimensionality $\dim{(D)}$, meaning the number of parameters.

\section{Software implementation} \label{sec:implementation}

\begin{figure}[!htbp]
    \centering
       \includegraphics[width=\columnwidth]{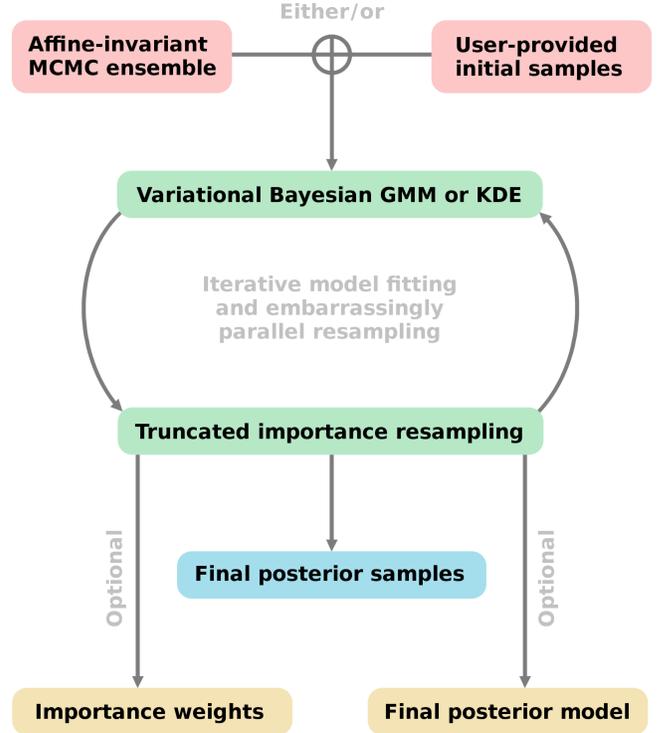}
   \caption{Schematic workflow of \texttt{Gaussbock}. Inputs are colored in red, iterative steps in green, primary outputs in blue, and optional outputs in yellow. Starting with an initial et of samples that roughly appproximates the posterior distribution, the method uses an iterative model-fitting and parallelized sampling-importance-resampling step using importance ratio truncation to evolve toward tighter fits for the true posterior. Depending on the dimensionality of the problem, a variational Bayesian Gaussian mixture model (GMM) or kernel density estimation (KDE) can be used. This iterative step is is repeated until convergence or a maximum number of iterations is reached. As indicated by the exclusive OR connection, the initial sample set can be user-provided or automatically inferred through a short-chained affine-invariant Markov chain Monte Carlo (MCMC) ensemble.}
    \label{fig:workflow}
\end{figure}

\begin{table*}[!htbp]
\renewcommand{\thetable}{\arabic{table}}
\centering
\caption{\texttt{Gaussbock} inputs. The table lists all 19 possible inputs that can be set by the user, as well as a short explanation for each input, with the first three being required. The remaining 16 optional inputs are marked with an asterisk before their name and are default values are based on the tests presented in this paper and should, as a result, generally achieve desirable performance for a wide array of problems reasonably similar to those described here.}
\begin{tabular}{lll}
\hline
\hline
Input & Explanation & Default \\
\hline
1. \texttt{parameter\_ranges} & The lower and upper limit for each parameter & \\
2. \texttt{posterior\_evaluation} & Evaluation function handle for the posterior & \\
3. \texttt{output\_samples} & Number of posterior samples that are required & \\
4. \texttt{*initial\_samples} & Choice of \texttt{emcee} or a provided start sample & [`automatic', $2 \cdot \dim{(D)} + 2$, $10^3$] \\
5. \texttt{*gaussbock\_iterations} & Maximum number of Gaussbock iterations & $10$ \\
6. \texttt{*convergence\_threshold} & Threshold for inter-iteration convergence checks & None \\
7. \texttt{*mixture\_samples} & Number of samples drawn for importance sampling & $10^4$ \\
8. \texttt{*em\_iterations} & Maximum number of EM iterations for the mixture & $10^3$\\
9. \texttt{*tolerance\_range} & The range for the shrinking convergence threshold & [$10^{-2}$, $10^{-7}$] \\
10. \texttt{*model\_components} & Maximum number of Gaussians fitted to samples & ceil$((2 / 3) \cdot \dim{(D)})$ \\
11. \texttt{*model\_covariance} & Type of covariance for the GMM fitting process & `full' \\
12. \texttt{*parameter\_init} & How to intialize model weights, means and covariances & `random' \\
13. \texttt{*model\_verbosity} & The amount of information printed during runtime & 1 \\
14. \texttt{*mpi\_parallelization} & Whether to parallelize Gaussbock using an MPI pool & False \\
15. \texttt{*processes} & Number of processes Gaussbock should parallelize over & 1 \\
16. \texttt{*weights\_and\_model} & Whether to return importance weights and the model & False \\
17. \texttt{*truncation\_alpha} & Truncation value for importance ratio reweighting & 2.0\\
18. \texttt{*model\_selection} & Type of model used for the fitting process & `gmm' if $\dim{(D)} > 2$, else `kde' \\
19. \texttt{*kde\_bandwidth} & Kernel bandwidth used when fitting via KDE & 0.5\\
\hline
\end{tabular}
\label{tab:params}
\end{table*}

In order to make this algorithm readily available, we have released a Python 3 package incorporating the complete \texttt{Gaussbock} algorithm. The package is installable via \texttt{pip} from the Python Package Index\footnote{\url{https://pypi.org}}, while documentation and source code are available in a public repository\footnote{\url{https://github.com/moews/gaussbock}}.

Figure~\ref{fig:workflow} shows the schematic workflow of \texttt{Gaussbock}, with a choice between an automated initial posterior approximation and a user-provided sample guess, as well as the option to return importance weights and the final fitted model. The automated initial approximation makes use of an affine-invariant MCMC ensemble, as introduced by \citet{Goodman2010}, through the package \texttt{emcee} by \citet{Foreman-Mackey2013} and with parameters like the number of walkers being automatically determined based on the required function inputs. The only required inputs to the tool's main function are the lower and upper limits for each parameter (`parameter\_ranges'), the handle of a function that accepts a point in the problem's parameter space and returns its log-posterior value (`posterior\_evaluation'), and the desired number of posterior samples to be returned (`output\_samples').

An overview of settable inputs is shown in Table~\ref{tab:params}. We strongly encourage users to provide parameter ranges that are scaled to the interval $[0, 1]$ when setting a threshold for the optional convergence determination (`convergence\_threshold') due to its mean variance-based functionality. When setting a convergence threshold, we recommend a value of $\sim 0.01 \cdot \dim(D)$ as a choice that, based on the tests performed in the course of this work, takes increased dimensionalities into account when using the built-in convergence criterion. The implementation uses \texttt{schwimmbad}, a library for parallel processing tools, to provide MPI parallelization on parallel computing architectures \citep{Price-Whelan2017}. The use of MPI can be activated with the optional boolean input (`mpi\_parallelization') being set to `True'. Alternatively, for running the algorithms across multiple cores locally, the optional input `processes' can be set to the number of desired cores to be used. The initial sample to start from can be provided by the user, for example through sampling a best-guess approximation or using the posterior from previous research (`initial\_samples').

An input of special importance is the ability to set the variable parameter for truncated importance sampling (`truncation\_alpha'), the ideal value of which can change based on the difficulty of the posterior approximation problem. By default, the recommended value of $2.0$ is used \citep{Ionides2008}. When dealing with, for example, high-dimensional truncated Gaussians or similarly hard-to-approximate shapes, a value of up to $3.0$ can enforce a stronger truncation to combat high-weight samples. Similarly, the truncation value can be set down to a minimum of $1.0$ for weaker importance weight truncation. Interlinked with this input are the dimensionality of the problem and number of samples drawn from a fitted model in each iteration (`mixture\_samples'), as a lower number of samples in a higher-dimensional parameter space increases the odds of importance weights with comparatively high values due to sparse samples. Time requirements and the number of available cores are the limiting factors for such considerations, which is discussed in the experiments in Section~\ref{sec:experiments}.

The algorithm's runtime can be further influenced by limiting the maximum number of Gaussians to be used for fitting a VBGMM during each iteration (`model\_components'). By default, this input is determined based on the number of parameters to be estimated, but user knowledge about the complexity of the target distribution can inform the requirement for lower or higher maximums. Low-dimensional problems with $\dim(D) < 3$ trigger the use of kernel density estimation (KDE) instead of a VBGMM by default, as this density estimation approach is quite powerful in such scenarios, but faces issues in higher-dimensional problems \citep{OBrien2016}. The use of KDE or a VBGMM can, however, be forced by the user by setting the respective optional input (`kde\_bandwidth') to either `kde' or `gmm'. The bandwidth used for the KDE functionality can be customized with an optional input (`kde\_bandwidth'). We advise the use of KDE for low-dimensional problems due to the ability to catch hard-to-approximate posteriors in combination with our iterative method, which we demonstrate in Section~\ref{sec:kde}.

\section{Experiments} \label{sec:experiments}

DES is an imaging survey that covers 5000 square degrees of the southern celestial hemisphere, operating a wide-field camera on the 4-meter Víctor M. Blanco Telescope located at the Cerro Tololo Inter-American Observatory \citep{Abbott2016a}. The survey probes cosmology using multiple different sources, including galaxy clustering and lensing, cluster counts, and supernova measurements. Preliminary constraints from DES Science Verification (SV) data are presented in \citet{Abbott2016b} and \citet{Kacprzak2016} while, more recently, results and data for DES Y1 observations are described by \citet{Abbott2018a} and have been made public\footnote{\url{https://des.ncsa.illinois.edu/releases/y1a1}}.

In this work, we use the Y1 weak lensing and galaxy clustering measurements as a test of \texttt{Gaussbock}. These measurements consist of a set of 2D two-point correlation functions of galaxy shape and position (``3x2pt'') in tomographic bins by redshift. These functions can be predicted from the cosmological matter power spectrum and redshift-distance relation, both of which are sensitive to the underlying cosmological parameters, and especially to the matter density fraction $\Omega_m$ and the variance of cosmic structure $\sigma_8$. DES analyses yield constraints on these parameters comparable to those obtained from the CMB with Planck \citep{Aghanim2018}. For our experiments, we use the baseline $\Lambda$CDM model with varied neutrino density as our test likelihood. The sampling methods used in the main DES analysis are discussed in \citet{Krause2018}; they use both the \texttt{emcee} affine-invariant sampler and the \texttt{MultiNest} nested sampling method, and found close agreement between the two methods.

In Section~\ref{sec:approximation}, we describe a fast-likelihood approximation of the DES Y1 posterior, followed by a performance test for \texttt{Gaussbock}. We explore scaling behavior of our implementation on the same approximation with experiments in Section~\ref{sec:scaling}. In Section~\ref{sec:des}, we run \texttt{Gaussbock} on the full DES Y1 posterior to test both the performance in real scenarios and the the ability to run fully parallelized via MPI on supercomputing facilities. Lastly, in Section~\ref{sec:kde} we test the behavior of the method on distributions with specific challenges and determine what types of failure modes it experiences.

\subsection{Approximating the Dark Energy Survey posterior}
\label{sec:approximation}

\begin{figure*}[!htbp]
    \centering
       \includegraphics[width=\textwidth]{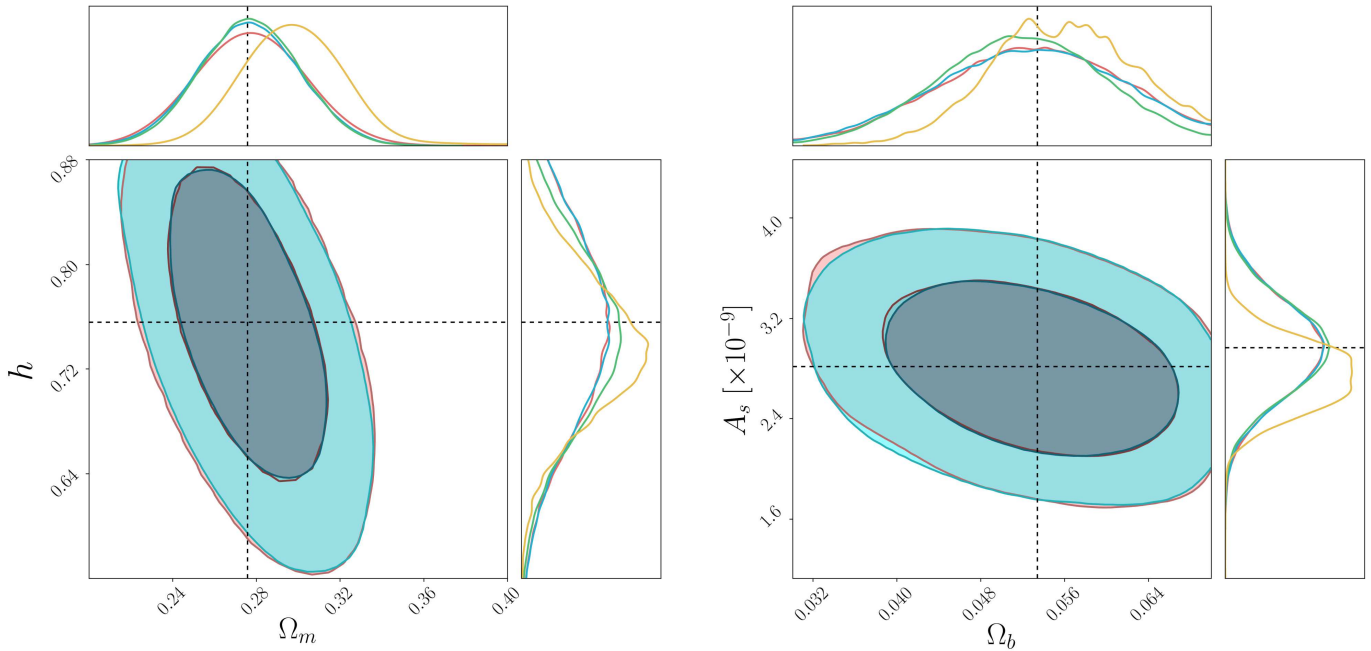}
   \caption{DES Y1 posterior approximation with \texttt{Gaussbock}. The left figure depicts the matter density parameter ($\Omega_m$) versus the Hubble constant ($H_0$), whereas the right figure shows the baryon density parameter ($\Omega_b$) versus the scalar amplitude of density fluctuations ($A_s$). Contours for the importance-weighted samples generated with \texttt{Gaussbock} are drawn in blue, with contours for an \texttt{emcee} chain with $5.4$ million samples across $54$ walkers drawn in red. Darker and lighter shaded contour areas depict the 68\% and 95\% credible intervals, respectively. In addition to the same color coding as used in the contour plots, one-dimensional subplots for each parameter also show the unweighted distribution of \texttt{Gaussbock} samples in green, and the initial guess from which \texttt{Gaussbock} starts, obtained through a short-chained \texttt{emcee} run with $1000$ steps per walker, in yellow. True means for DES Y1 data are indicated with dashed black lines to demonstrate the correct centering of both the fast approximation we employ in the experiment and the \texttt{Gaussbock} outputs.}
    \label{fig:contours}
\end{figure*}

The real DES Y1 likelihood is slow to evaluate, with durations per likelihood that make serial algorithms non-viable, as in \citet{Wilkinson2005}. In order to enable experiments that target controlled assessment and scaling behavior, we use an approximation to the DES Y1 posterior with a multivariate truncated Gaussian distribution, for which we employ the mean and covariance values for 26 cosmological and nuisance parameters, as well as their limits from the respective DES data release. This approach results in an extremely fast parameter set evaluation based on a DES Y1 approximation suitable for our purposes. A perfectly Gaussian approximation to the posterior would be an artificially easy test of a model that fits Gaussians; our posterior is truncated within a few sigma of the peak in many of its parameters, and thus provides a reasonable challenge.

As discussed in Section~\ref{sec:implementation}, we use an increased truncation value for the SIR step of \texttt{Gaussbock}, which we set to $3.0$, and a convergence threshold of $0.01 \cdot 26 = 0.26$ that follows the previously outlined best-practice guidelines and triggers the use of the built-in convergence determination. The number of samples per iteration is set to $15000$, with the reasoning behind this choice further outlined in Section~\ref{sec:scaling}. As we want to weight the returned posterior samples with their importance weight, we activate the additional return of the final model and importance weights. Apart from these settings, we use the default behavior of \texttt{Gaussbock} by not providing other optional inputs. Table~\ref{tab:limits} shows the lower and upper limits for cosmological and nuisance parameters employed in our approximation.

\begin{table}[!htbp]
\renewcommand{\thetable}{\arabic{table}}
\centering
\caption{Cosmological and nuisance parameter limits for a fast approximation of the DES Y1 posterior. The lower and upper limits shown as open intervals closely follow prior distribution features previously used by DES for data from the first year of observations \citep{Abbott2018a}.}
\begin{tabular}{lll}
\hline
\hline
Category & Parameter & Interval \\
\hline
Cosmology & $\Omega_m$ & $[0.1, 0.9]$ \\
 & $H_0$ & $[0.55, 0.9]$ \\
 & $\Omega_b$ & $[3 \cdot 10^{-2}, 7 \cdot 10^{-2}]$ \\
 & $n_s$ & $[0.87, 1.07]$ \\
 & $A_s$ & $[5 \cdot 10^{-10}, 5 \cdot 10^{-9}]$ \\
 & $\omega_\nu$ & $[6 \cdot 10^{-4}, 10^{-2}]$ \\
\hline
Lens galaxy bias & $b_1, \dots, b_5$ & $[0.8, 3.0]$\\
\hline
Shear calibration & $m_1, \dots, m_4$ & $[-0.1, 0.1]$\\
\hline
Intr. alignment & $A_{\mathrm{IA}}$ & $[-5.0, 5.0]$ \\
 & $\mu_{\mathrm{IA}}$ & $[-5.0, 5.0]$ \\
\hline
Source photo-$z$ & $\Delta z_s^1, \dots, \Delta z_s^4$ & $[-0.1, 0.1]$ \\
\hline
Lens photo-$z$ & $\Delta z_l^1, \dots, \Delta z_l^5$ & $[-5 \cdot 10^{-2}, 5 \cdot 10^{-2}]$ \\
\hline
\end{tabular}
\label{tab:limits}
\end{table}

\begin{figure*}[!htbp]
    \centering
       \includegraphics[width=\textwidth]{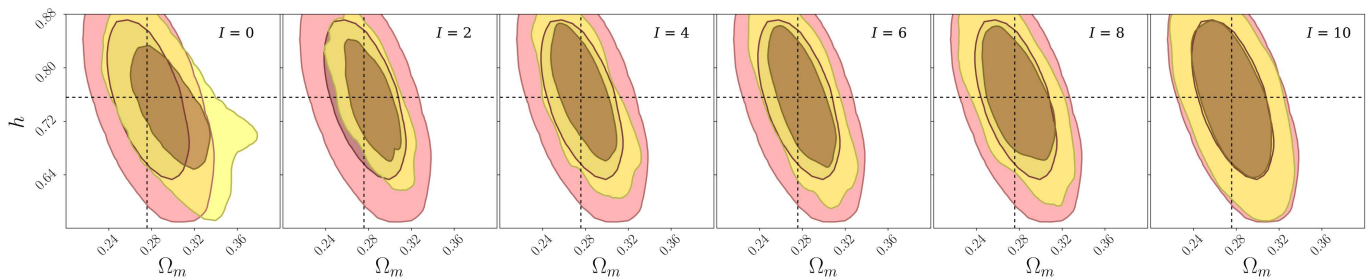}
   \caption{Gradual improvement of contours across \texttt{Gaussbock} iterations. The figure depicts, in yellow, the importance-weighted posterior approximations for the matter density parameter ($\Omega_m$) versus the Hubble constant ($H_0$). Each panel indicates the respective number of iterations $I$ in the upper right corner, for iteration numbers from the the set $\{0, 2, \dots 10\}$ to cover easily visible morphing behavior before fine-tuning takes place. Contours for an \texttt{emcee} chain with $5.4$ million samples across $54$ walkers are drawn in red to serve as a target distribution and orientation point across panels. Darker and lighter shaded contour areas depict the 68\% and 95\% credible intervals, respectively. On the far left, at $I = 0$, the posterior approximation corresponds to the initial sample guess. True means for DES Y1 data are indicated with dashed black lines.}
    \label{fig:morphing}
\end{figure*}

The results of this experiment are shown in Figure~\ref{fig:contours} and demonstrate the abilily of \texttt{Gaussbock} to recover correct constraints. Starting from a short and unconverged \texttt{emcee} chain, for which distributions are shown in yellow, the importance-weighted posterior samples marked in blue closely match the long-run \texttt{emcee} samples highlighted in red. The achieved level of agreement is good enough to make posterior contours and distributions for the target distribution and the importance-weighted samples hard to separate by eye. While the distributions for unweighted posterior samples in green show a good agreement with the long-run samples, weighting the output samples with the optionally returned importance weights pushes the sample distributions further toward to target posterior, thus validating the additionally provided functionality related to KDE for low-dimensional parameter estimation. While this experiment is based on an approximation of the full DES Y1 posterior, it offers a suitable testbed to prepare for the full-scale run described in Section~\ref{sec:des}. 

Another factor of interest is the iterative behavior of our algorithm, as \texttt{Gaussbock} is supposed to continuously improve the agreement of its internally generated samples with the true posterior distribution. In Figure~\ref{fig:morphing}, we illustrate this behavior, showing the gradual improvement of the constraints. The plots depict the morphing and shifting behavior of \texttt{Gaussbock} samples for the number of iterations as even integers in the interval [0, 10]. The cosmological parameters chosen for this experiment are the same as in the left-hand panel of Figure~\ref{fig:contours}.

The evolution across the different panels showcase the algorithm's ability to start from a very rough sample guess and gradually move toward the target distribution. The latter is closely approximated by an extremely long \texttt{emcee} chain as an ideal sample. As demonstrated through this visualization, the algorithm first shifts generated samples toward the true mean with a lower-variance distribution, followed by incrementally spreading out to create a close fit to the target distribution.

\subsection{Exploration of scaling behavior} \label{sec:scaling}

In algorithms designed for the use with highly parallelized architectures, as well as in approaches for high-dimensional estimation problems, the question of how the algorithm in questions scales for different factors is important. For this reason, we now explore the scaling behavior of our algorithm. We quantify the time to convergence using the criterion introduced in Section~\ref{sec:estimation}, measured on the fast DES Y1 approximation covered in Section~\ref{sec:approximation}.

\begin{figure*}[!htbp]
    \centering
       \includegraphics[width=\textwidth]{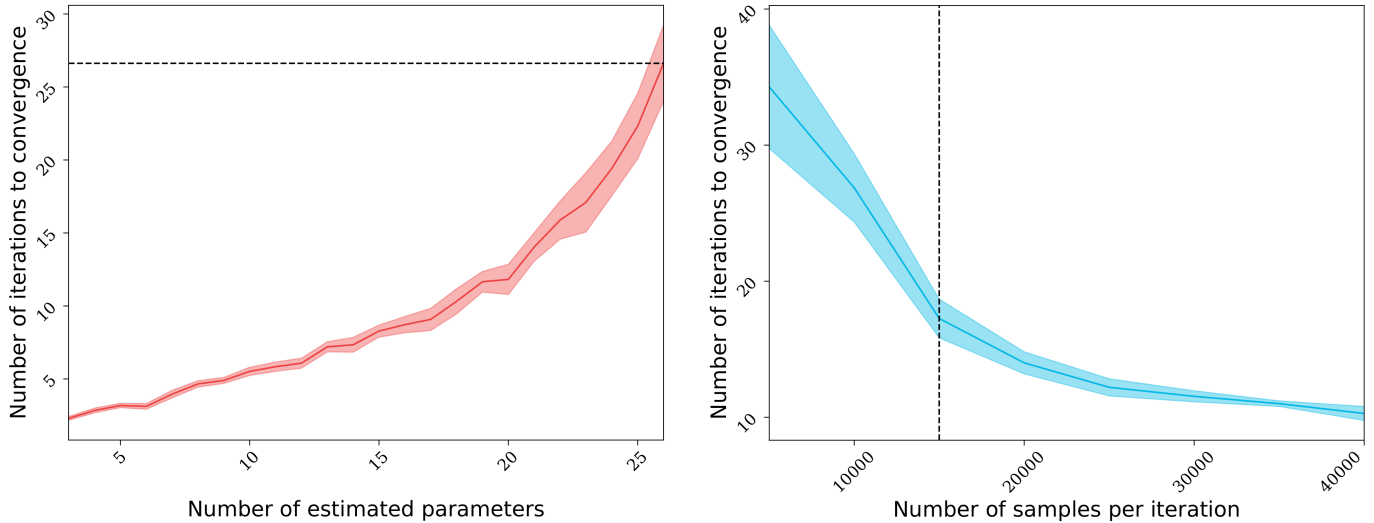}
   \caption{Relationship between time to convergence, dimensionality, and the number of samples per iteration for \texttt{Gaussbock}. The left panel shows the number of iterations needed to achieve convergence, as a function of the dimensionality of the problem. The dashed black line indicates the mean number of iterations (26.6) needed for the full 26D DES Y1 parameter set. The right panel shows the number of iterations before convergence, as a function of the number of importance samples taken at each iteration, in steps of 5000. The dashed line  marks the `elbow criterion' for the trade-off in terms of time requirements from iterations and sample size, at 15000 samples. In both panels, the central line shows the mean and the shaded band the 95\% confidence intervals over 50 simulations per point.}
    \label{fig:convergence_iterations}
\end{figure*}

Higher-dimensional problems can, in general, be assumed to lead to a greater complexity of the estimation procedure, forcing \texttt{Gaussbock} to morph and shift the distribution in each iteration across more dimensions. We test our implementation for dimensionalities $3 \leq \dim(D) \leq 26$, up to the full set of cosmological and nuisance parameters in our DES Y1 approximation. We perform this parameter estimation 50 times for each number of dimensions to create confidence intervals, with the respective subset of parameters being randomly selected. In each case, we use the convergence threshold $0.01 \cdot \dim{(D)}$. The left panel of Figure~\ref{fig:convergence_iterations} plots the number of iterations required to reach convergence versus the number of estimated parameters, showing the rise with problems of increased dimensionality.

The 95\% confidence intervals around the average number of iterations to convergence highlight the larger variance with increasing numbers of parameters. The average number of 26.6 iterations for estimating the full set of 26 parameters provides an indicator for the full DES Y1 posterior computation in Section~\ref{sec:des}.

The second question in terms of scaling behavior targets the embarrassingly parallel part of our algorithm, as we can vary the number of samples drawn at each iteration. Although the ability to parallelize across large numbers of cores is one of the strengths of \texttt{Gaussbock}, and while access to parallel computing architectures is wide-spread in modern cosmology, the number of available cores for a given task still faces upper limits. As described in Section~\ref{sec:implementation}, a higher number of samples drawn from a given iteration's fitted model is generally preferable, which translates to a preference for a higher number of cores due to the subsequent parallelization of the truncated SIR step. This poses the question of the scaling behavior of this benefit, as the required number of iterations to convergence is expected to decrease with a higher number of samples per iteration.

The right panel of Figure~\ref{fig:convergence_iterations} shows the scaling behavior of the required number of iterations to convergence versus the number of samples drawn from the fitted model during each iteration. We perform 50 \texttt{Gaussbock} runs per number of samples to create confidence intervals, in the interval [5000, 40000] and in steps of 5000.

Let $I$ be the number of required iterations to convergence, $C$ the number of available cores, and $S$ the number of used samples per iteration. Then the total number of posterior value calculations per core over the course of a \texttt{Gaussbock} run is $I \cdot S \cdot C^{-1}$. Increasing numbers of samples constrain the variance of required iterations, and the dashed black line in the right panel of Figure~\ref{fig:convergence_iterations} indicates an optimal trade-off (in terms of total core time) between the two variables as $\min(I \cdot S)$ at $S = 15000$ for the number of samples, which informs our input choices in Section~\ref{sec:approximation}. This visualization also bears resemblance to the `elbow criterion' in cluster analysis, which determines the optimal number of clusters by plotting that number against the explained variance \citep{Thorndike1953}.

\begin{figure}[!htbp]
    \centering
       \includegraphics[width=\columnwidth]{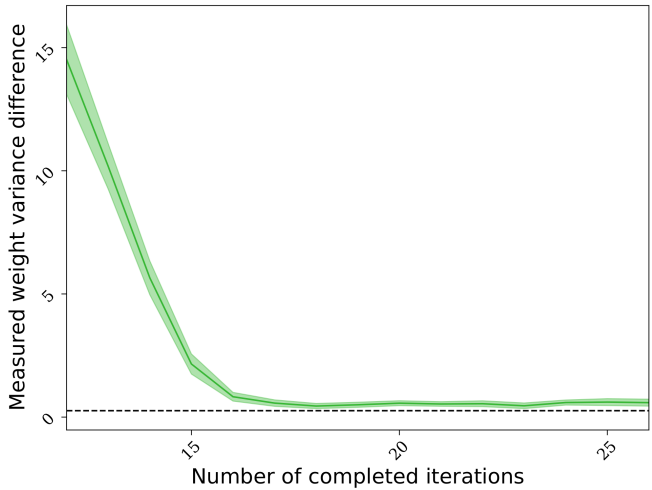}
   \caption{Convergence behavior of \texttt{Gaussbock} for the number of completed iterations in approximated 26D DES Y1 analyses. The figure shows the inter-iteration change in variances of the logarithmic weights, used as a convergence criterion, with the dashed line marking the default convergence threshold for this problem. The mean value over 50 runs is shown as the central line, and the shaded band shows the 95\% confidence interval.
   }
    \label{fig:completed_iterations}
\end{figure}

Lastly, we investigate the convergence behavior of \texttt{Gaussbock} as a follow-up to Figure~\ref{fig:morphing}, to ensure that both the convergence check itself and the recommended calculation of a convergence threshold behave as intended. The algorithm is run on the same parameter estimation problem as in Section~\ref{sec:approximation}, for a total of 27 iterations to cover the previously computed mean number of iterations to convergence of 26.6. As for previous tests, we run this experiment 50 separate times to generate 95\% confidence intervals. The results are shown in Figure~\ref{fig:completed_iterations}, starting after the first 10 iterations to cover fine-tuning behavior after the initial morphing and shifting explored in Figure~\ref{fig:morphing}, and with the dashed black line indicating the convergence threshold set to $0.01 \cdot \dim(D) = 0.26$. The figure, showing a remarkably consistent and well-constrained behavior, demonstrates both convergence behavior for the threshold calculation and narrow confidence intervals for multiple experiments.

\subsection{The full Dark Energy Survey posterior} \label{sec:des}

In order to expose our method to a fully realistic experiment without approximations, we apply \texttt{Gaussbock} to the full DES posterior from the DES Y1 experiments and data release \citep{Abbott2018a, Abbott2018b, Krause2018}. We use the public \texttt{CosmoSIS} implementation of the public Y1 likelihood, which includes \texttt{CAMB} as descibed by \citet{Lewis2002} and \citet{Howlett2012}, and \texttt{Halofit} as introduced in \citet{Smith2003} and \citet{Takahashi2012} to compute distances and matter power spectra, \texttt{CosmoSIS}-specific modules for the Limber integral and other intermediate steps, and \texttt{Nicaea}\footnote{\url{http://www.cosmostat.org/software/nicaea}} for the computation of real-space correlations from Fourier space values \citep{Kilbinger2009}. Since the public implementation of the Y1 likelihood differs very slightly from the released chains, we rerun the model referred to as \texttt{d\_l3} in the public DES Y1 chains using \texttt{MultiNest} for an identical comparison.
The experiment starts with the same initial sample guess via a short-chained \texttt{emcee} run that we use for our fast approximation of the DES Y1 posterior in Section~\ref{sec:approximation}, demonstrating that our approach is able to start from approximative guesses that only partially fall within the vicinity of the target posterior and are not necessarily based on calculations using the actual target in question.

Making use of \texttt{Gaussbock}'s innate embarrassing parallelism, we run this experiment on supercomputing facilities of the National Energy and Scientific Research Computing Center (NERSC)\footnote{\url{https://www.nersc.gov}} \citep{He2018}. We run on 32 nodes of the Cori computer, for a total of 1024 cores and 2048 threads. The results below were generated in approximately two hours in this configuration, showcasing the total runtime advantage of our approach. With the runtime scaling being inversely linear with the number of cores, up to the number of samples used per iteration due to the model-fitting process not requiring a lot of time, up to 15000 cores can be used in an idealized scenario for our experimental setup to gain a further order-of-magnitude reduction. In order to make use of existing posterior implementations, we employ \texttt{CosmoSIS} to use \texttt{Gaussbock} with the DES Y1 posterior \citep{Zuntz2015}.

\begin{figure*}[!htbp]
    \centering
       \includegraphics[width=\textwidth]{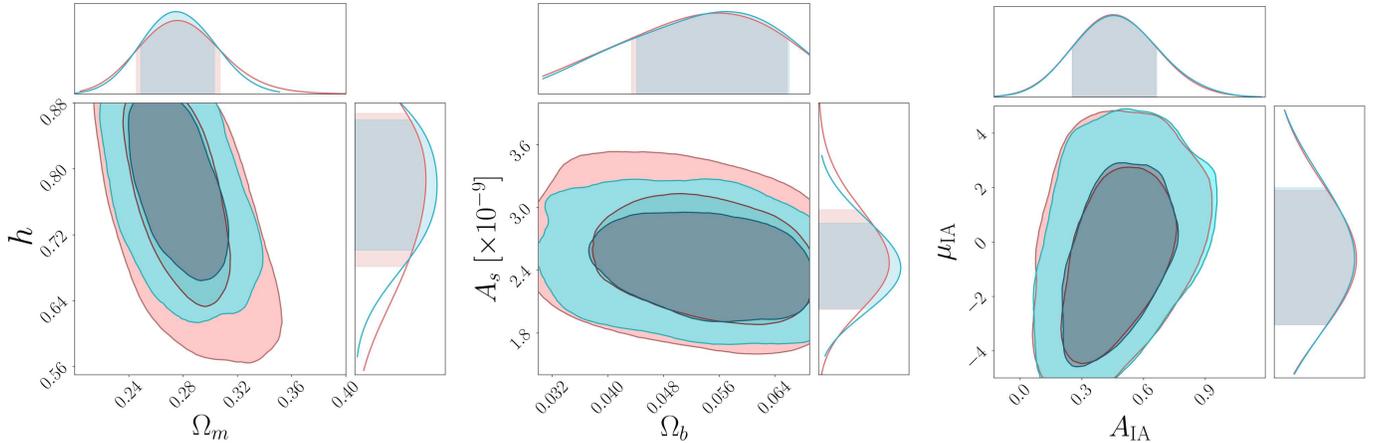}
   \caption{DES Y1 posteriors with \texttt{Gaussbock}. The left panel depicts the matter density parameter ($\Omega_m$) versus the Hubble constant ($H_0$), the middle figure shows the baryon density parameter ($\Omega_b$) versus the scalar amplitude of density fluctuations ($A_s$), and the right figure shows the two intrinsic alignment parameters ($A_{\mathrm{IA}}$, $\mu_{\mathrm{IA}}$). Contours for the importance-weighted samples generated with \texttt{Gaussbock} are drawn in blue, with contours for the original nested sampling implementation as used by DES drawn in red. Darker and lighter shaded contour areas depict the 68\% and 95\% credible intervals, respectively, with the same levels shaded in the histograms.}
    \label{fig:full}
\end{figure*}

Table~\ref{tab:results} lists the cosmological parameters as estimated by both \texttt{Gaussbock} and its comparison baseline, meaning the fiducial \texttt{MultiNest} run, demonstrating a satisfactory level of agreement for both means and credible intervals. In addition to the cosmological parameters shown in the experiments for Figures~\ref{fig:contours} and \ref{fig:full}, the table also includes the scalar spectral index $n_s$ and the massive neutrino density $\omega_\nu$, covering the full set of cosmological parameters previously listed in Table~\ref{tab:limits}.

\begin{table}[!htbp]
\renewcommand{\thetable}{\arabic{table}}
\centering
\caption{Cosmological parameters for DES Y1 data. The table shows figures of merit for common cosmological parameters used in the original DES Y1 experiments, with the latter's implementation of \texttt{MultiNest} and, for comparison, the results for a highly parallel \texttt{Gaussbock} run.}
\begin{tabular}{lll}
\hline
\hline
Parameter & \texttt{MultiNest} & \texttt{Gaussbock} \\
\hline
\vspace{5pt}
$\Omega_m$ & $0.276^{+0.031}_{-0.031}$ & $0.275^{+0.029}_{-0.026}$ \\
\vspace{5pt}
$H_0$ & $0.787^{+0.080}_{-0.106}$ & $0.781^{+0.078}_{-0.080}$ \\
\vspace{5pt}
$\Omega_b$ & $0.056^{+0.010}_{-0.012}$ & $0.057^{+0.009}_{-0.013}$ \\
\vspace{5pt}
$n_s$ & $1.020^{+0.043}_{-0.064}$ & $1.013^{+0.043}_{-0.065}$ \\
\vspace{5pt}
$A_s$ & $2.470^{+0.510}_{-0.440} \ \times 10^{-9}$ & $2.430^{+0.420}_{-0.400} \ \times 10^{-9}$\\
\vspace{5pt}
$\omega_\nu$ & $5.100^{+2.900}_{-2.800} \ \times 10^{-3}$ & $5.000^{+3.000}_{-2.800} \ \times 10^{-3}$ \\
\hline
\end{tabular}
\label{tab:results}
\end{table}

Figure~\ref{fig:full} shows the posterior contours for both the \texttt{d\_l3} rerun with \texttt{MultiNest} and the \texttt{Gaussbock} result in red and blue, respectively. Both matter and baryon density parameters, $\Omega_m$ and $\Omega_b$, are shown to match the baseline computation well, whereas the Hubble parameter $H_0$ and scalar amplitude of density fluctuations $A_s$ are in reasonable agreement, but do not correctly recover the tails of the posterior distribution. An exploration of the 26-dimensional approximation shows that \texttt{Gaussbock} accurately models the parameters which are well-constrained, but fails to recover the tails on unconstrained parameters like $H_0$ and $A_a$ that have very broad intervals, as listed in Table~\ref{tab:limits}. Where possible, it might help to provide narrower constraints for such parameters. In addition, Figure~\ref{fig:full} shows the joint posterior of the two intrinsic alignment parameters,  $A_{\mathrm{IA}}$ and $\mu_{\mathrm{IA}}$ in the right panel.

The results demonstrate the ability of \texttt{Gaussbock} to recover non-Gaussian shapes of correlated parameters to a high degree of accuracy, as can be seen in the 2D posterior shapes for the fiducial \texttt{MultiNest} and \texttt{Gaussbock} runs, as well as in the agreement between histograms in the figure. The effective sample size for this \texttt{Gaussbock} run is $N_\mathrm{eff}=2104$, compared to $N_\mathrm{eff}=4316$ for the  original \texttt{MultiNest} chain, although with a smaller overall runtime for our algorithm.

While the results are not in near-perfect agreement, as is the case for the fast truncated Gaussian approximation in Section~\ref{sec:approximation}, a trade-off between considerably reduced runtime and accuracy is to be expected analogous to the No Free Lunch Theorem in optimization \citep{Wolpert1997}. The described experiment on the full DES Y1 posterior makes use of \texttt{Gaussbock}'s adaptive default behavior and, for the number of samples per iteration, is based on our fast approximation, so fine-tuning to a specific application case can be expected to further improve the performance of the algorithm. Other reasons for the results not showing the same goodness of fit for all parameters, as observed in Section~\ref{sec:approximation}, are a diminished smoothness of posteriors and less Gaussian tails, which we discuss in Section~\ref{sec:results}.

\subsection{Stress tests on additional distributions} \label{sec:kde}
In this subsection we run \texttt{Gaussbock} on distributions with  more challenging features to determine when it starts to fail. As outlined in Section~\ref{sec:implementation}, KDE is a powerful density estimation technique, but faces issues in higher-dimensional problems \citep{OBrien2016}. In this experiment, we exemplify the built-in default to use KDE for problems in which $\dim(D) \leq 2$, allowing \texttt{Gaussbock} to make use of the method most suitable to a given problem. For this purpose, we construct a posterior of three approximately equilateral triangles with a flat posterior surface, meaning that posterior values are uniform across the triangle shapes. Due to the convergence criterion of \texttt{Gaussbock}, which we discuss in Section~\ref{sec:estimation}, being geared toward the use of a VBGMM as its primary application in high-dimensional setting, we set the number of iterations to 20. We let the initial sample guess be generated automatically with the same number as for previous experiments in Section~\ref{sec:approximation}, and let \texttt{Gaussbock} use its default behavior for optional inputs.

\begin{figure}[!htbp]
    \centering
       \includegraphics[width=\columnwidth]{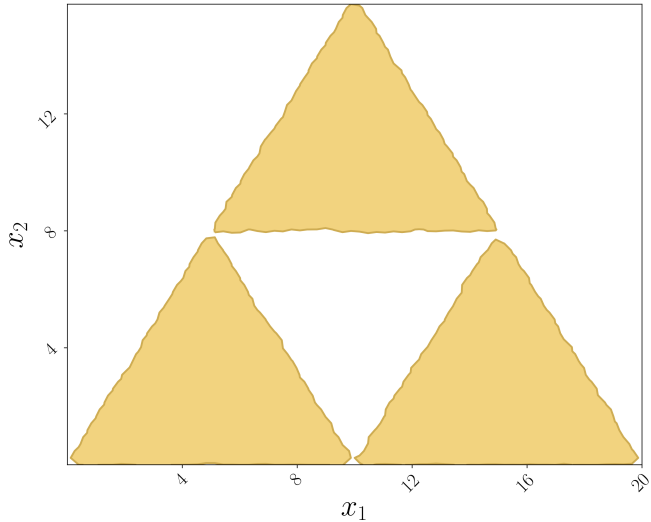}
   \caption{Approximation of a hard-to-estimate posterior with \texttt{Gaussbock}. The two-dimensional posterior distribution features uniform values across the surface of three triangles. With a completely flat distribution of the posterior shape, the importance-weighted sample contours in the plot show the 95\% credible interval for the generated samples.}
    \label{fig:lowdim}
\end{figure}

The results of this low-dimensional parameter estimation experiment is shown in Figure~\ref{fig:lowdim}, with 95\% credible intervals for the flat-surface posterior demonstrating the ability of \texttt{Gaussbock} to approximate complex shapes with pronounced edges and corners. The three separate triangles are clearly reconstructed through the importance-weighted samples generated by the algorithm, validating its integrated KDE functionality for low-dimensional estimation problems.

\begin{figure}[!htbp]
    \centering
       \includegraphics[width=\columnwidth]{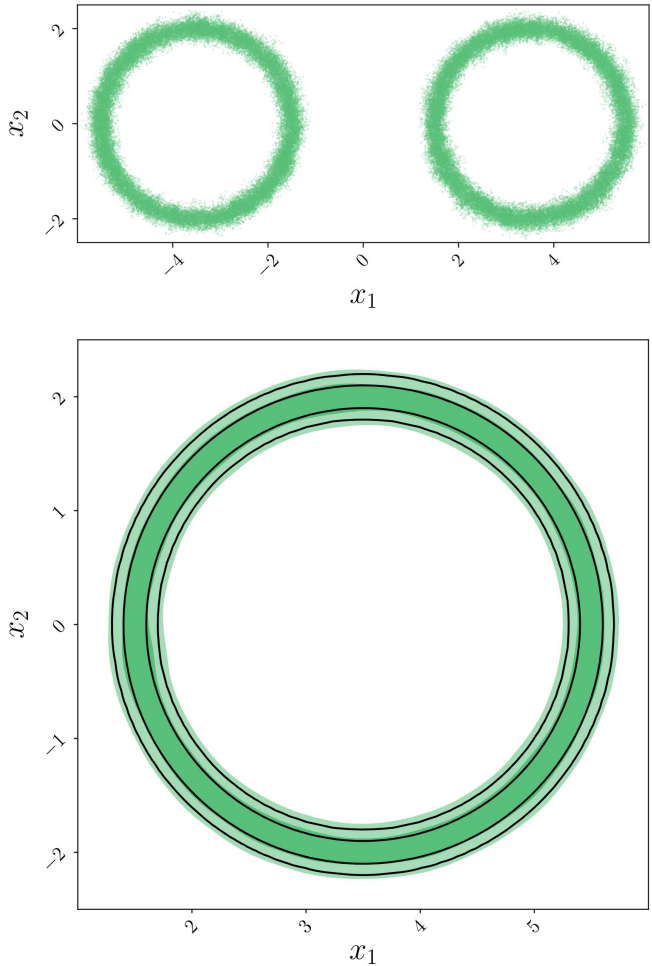}
   \caption{Samples from a 2D Gaussian shell distribution. The upper panel shows a scatter plots of the resulting \texttt{Gaussbock} samples, while the lower panel zooms in on one of the two shells. For the latter, we show inner 68\% and outer 95\% contours from a brute-force grid evaluation in black, and KDE on \texttt{Gaussbock} samples as blue-shaded regions, with darker and lighter shaded contour areas depicting the 68\% and 95\% credible intervals, respectively. At higher dimensions, \texttt{Gaussbock} fails on such distributions.}
    \label{fig:shell}
\end{figure}

Next, we consider  similar stress tests  based on those described in \citet{Hobson2008} and \citet{Feroz2009}.  First, we test with a posterior in the form of a double Gaussian shell, as described in \citet{Allanach2008},
\begin{equation}
    {\cal L}(\theta) = C(\mathbf{\theta}; \mathbf{c_1}, w, r) +  C(\theta; \mathbf{c_2}, w, r),
\end{equation}
where
\begin{equation}
    C(\theta; \mathbf{c}, w, r) = {\cal N}(\left| \mathbf{\theta} - \mathbf{c} \right|; r, w^2).
\end{equation}
At low dimensions, Gaussbock can sample effectively from such a distribution; the results from a 2D example with $w = 0.1$ and $r = 2$ are shown in Figure~\ref{fig:shell}.  The samples correctly trace the distribution, with a close-to-ideal match between the brute-force percentiles and the fraction of samples inside them. At moderate dimensions, from around 5D, Gaussbock fails on the sharp edges in this distribution, as the required number of Gaussians to capture the full shape becomes too high.

\begin{figure}[!htbp]
    \centering
       \includegraphics[width=\columnwidth]{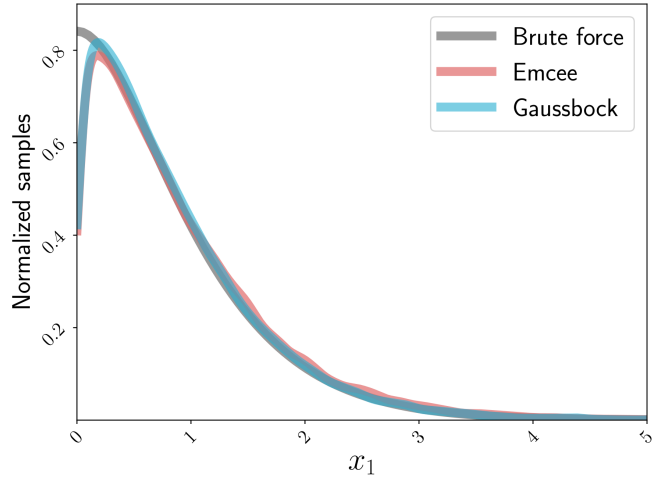}
   \caption{Sampling behavior of \texttt{Gaussbock} on the distribution in Equation~\ref{eq:expon}, with a sharp boundary in 4D, compared to a long-chained \texttt{emcee} run and a brute-force evaluation. Both samplers underestimate the PDF near the edge, although \texttt{Gaussbock} maintains a slightly smoother adherence to the true distribution otherwise.}
    \label{fig:expon}
\end{figure}

Next, we consider sharp edges that are poorly fit by Gaussian mixtures as another possible failure cases. We sample from
\begin{equation}
    {\cal L} = 
        \begin{cases}
      \exp{-2 \cdot |\mathbf{x}|}, & \text{if}\ \forall x \in \mathbf{x} :  x > 0 \\
      0, & \text{else}
    \end{cases}
    \label{eq:expon}
\end{equation}
using a 4D example. This form has a sharp edge at $\mathbf{x}_i = 0$ in each dimension. Figure~\ref{fig:expon} shows the 1D distribution of one of the four parameters, as sampled using \texttt{Gaussbock}, \texttt{emcee}, and a brute-force evaluation. Both samplers undersample at this boundary\footnote{Many sampling methods based on Markov chains can suffer from repulsive effects at sharp edges of distributions, since proposals to points near the boundary can only happen from one direction; a variety of methods have been used to correct for this behavior  \citep{Ahmadian2011}.}, and this effect will worsen for \texttt{Gaussbock} at higher dimension.

\begin{figure}[!htbp]
    \centering
       \includegraphics[width=\columnwidth]{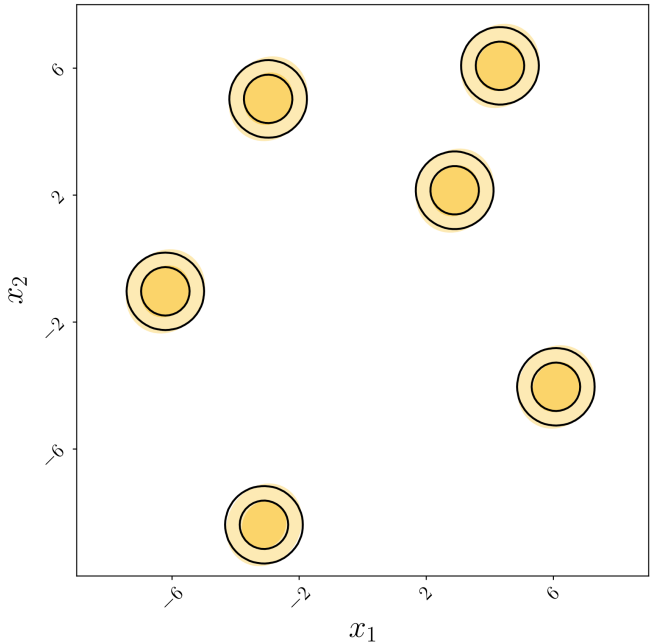}
   \caption{A 2D projection of a six-dimensional distribution with six Latin hypercube-located Gaussian modes. We show a KDE on \texttt{Gaussbock} samples as yellow-shaded regions, with darker and lighter shaded contour areas depicting the 68\% and 95\% credible intervals, respectively. The algorithm typically finds all the modes up to about 6D, and then begins to miss them at higher dimensions due to the difficulty of catching them in the initial sample generation..}
    \label{fig:multi}
\end{figure}

Finally, as a multimodal example, we consider sets of identical Gaussians, with centers arranged in a Latin hypercube formation so that they do not overlap in any dimension. The algorithm, starting from a random scattering throughout the space, finds all the modes for dimensions up to about six, as shown in Figure~\ref{fig:multi}. At higher dimensions, the algorithm often misses some of the modes; this is an important failure case that is based in the reliance on an initial sample provided by either the built-in affine-invariant MCMC sampler or a method of choice. If the latter fails to catch at least part of some modes, the algorithm is unlikely to recover them.

It should be noted that most distributions found in the additional tests of this section are not usually found in the intended field of application, cosmological parameter estimation, but serve as a demonstration of the method's capabilities for classical tests found in the statistical literature, and could be of use in other application areas.

\section{Discussion} \label{sec:results}

The primary advantage of our approach is the considerable reduction in the required runtime, given a large-enough number of cores available for parallelization. This strength offers a way to tackle increasing complexities in cosmological parameter estimation for current and upcoming surveys such as LSST and Euclid \citep{Amendola2018}. Since cosmological parameter estimation efforts rely on computationally costly posterior evaluations, the embarrassing parallelization of their calculation allows for an immense speed-up in comparison to standard MCMC approaches. This reduction in total runtime comes, however, at the cost of an increase in the required core time, meaning the number of computing hours necessary to achieve suitable results. For this reason, and assuming a sufficiently costly posterior evaluation, making use of \texttt{Gaussbock}'s parallelization capabilities is a requirement rather than an optional feature, as demonstrated in Section~\ref{sec:des}.

A direct comparison to MCMC methods is a double-edged sword in that such methods, run for a very large number of steps, provide a close fit to the true posterior. The downside of MCMC approaches is that they tend to not scale well with the number of dimensions, and that they are only parallelizable over the number of walkers. This means that computationally expensive likelihoods provide an obstacle to implementations such as \texttt{emcee} \citep{Foreman-Mackey2013}. While nested sampling methods circumvent this restriction by requiring fewer posterior evaluations, they rely on assumptions about perfect and independent samples and can sometimes underestimate an asymptotically Gaussian sampling error. In many cases, though, they can be highly effective, for both posterior and evidence estimation, depending on the problem at hand \citep{Chopin2010}.

As mentioned in Section~\ref{sec:des}, posteriors based on real-world survey data may have a less smooth posterior surface, which can hamper the effectiveness of the truncated SIR step used in our approach. Adjusting the `truncation\_alpha' input can alleviate this issue for isolated samples with higher posterior values, although a more effective solution is to increase the number of samples drawn from the posterior approximation of a given iteration of the algorithm. This approach does, in turn, require either a correspondingly larger number of cores or additional runtime. Alternatively, the initial sample guess to which the first-iteration model is fitted can be based on a longer-chain \texttt{emcee} run. As a result, this approach offers a better approximation of the posterior to start from, as it more closely resembles the target distribution and leads to broader coverage of relevant areas. We hope that the presented work will lead to further investigations of this and related parallelized iterative approaches to parameter estimation, alleviating the issues arising from increased computational demands in inference based on modern surveys.

Apart from cases with sufficiently smooth posteriors and well-constrained parameters, \texttt{Gaussbock} also offers a way to quickly approximate a posterior to reasonable degrees. For this purpose, we recommend using either uniform-random samples from an $n$-sphere scaled to the admissible ranges or, if feasible, samples from a better-suited distribution like an $n$-dimensional Gaussian to provide an initial sample guess covering the posterior area. The reason for such approaches is the elimination of the need for computationally more expensive sample guess generators such as short-chained \texttt{emcee} runs, which require costly evaluations of the posterior. While short chains are fast in comparison to exhaustive runs of MCMC methods, runtimes should be kept to a minimum for fast approximations in order to provide an edge in speed over alternative approaches.

An additional use case pertains to lower-dimensional problems, or scenarios with posterior evaluations that are sufficiently cheap to compute, and offers a way to achieve very tight fits to posteriors that are hard to approximate and feature clean cuts, with an example given in Section~\ref{sec:kde} and one commonly-encountered example of such posterior shapes being truncated Gaussians. The suitability for the latter type also extends to higher dimensions, as we demonstrate with the truncated Gaussian approximation of the 26-dimensional DES Y1 posterior in Section~\ref{sec:approximation}. For the latter, as described in Section~\ref{sec:des}, an important finding is that \texttt{Gaussbock} accurately models well-constrained parameters, but can have trouble to recover the tails on unconstrained parameters perfectly. For that reason, setting sensible parameter constraints as one of the three required inputs to the implementation is strongly advised.

Unlike in most MCMC methods, the final mixture model is an optional output of our implementation, which can be saved and used again at a later point. It can act as an approximate but analytic description of the posterior, allowing, for example, the subsequent drawing of an arbitrary number of samples for which importance weights can be calculated and which can be easily disseminated. In this context, our approach offers a way to easily exchange and compare posterior approximations based on different datasets, with mixture models whose components can be combined.

For common problems faced by contemporary research in cosmology, \texttt{Gaussbock} offers a considerable speed-up. This is especially relevant for upcoming missions with larger numbers of parameters, for which our approach provides a way to quickly compute high-fidelity posterior approximations and the underlying mixture model. While, in this work, we use a wrapper to run \texttt{Gaussbock} through \texttt{CosmoSIS} on NERSC facilities, a complete integration into \texttt{CosmoSIS} will further enhance the ease of access to our methodology. Regarding the scaling behavior tested in Section~\ref{sec:scaling}, a higher number of dimensions leads to a higher number of iterations to reach convergence, as demonstrated in Figure~\ref{fig:convergence_iterations}. \texttt{Gaussbock} also benefits from an as-close-as-possible fit to the true posterior for the initial sample to start from. In cases in which such a sample guess is available, it lends an advantage to the method's performance when compared to using the built-in affine-invariant MCMC sampler. Notably, the ability to feed an arbitrary set of initial samples into the tool also means that \texttt{Gaussbock} can be combined with any sampling algorithm to create such an initial sample, allowing users to employ cutting-edge methods of their choice to make full use of the current statistical literature and personal preferences.

In terms of its internal functionality, our approach inherently lends itself to combating issues with defaulting cores, as the failure or a subset of processes to return importance values can be safely ignored. The respective parameter sets can simply be omitted from the set of samples used to approximate the posterior in a given iteration, using the large-enough amount of remaining parameter sets to fit the model in a given iteration. While the capability to do so is not part of our implementation and is primarily of interest for large-scale cloud computing, our code easily lends itself to being extended toward this safety redundancy.

\section{Conclusion} \label{sec:conclusion}

In this paper, we introduce and apply \texttt{Gaussbock}, a novel approach to cosmological parameter estimation that makes use of recent advances in machine learning and statistics. By coupling variational Bayesian GMMs with a truncation-based extension of importance sampling in an iterative approach with a convergence criterion, our method offers an embarrassingly parallel way to achieve high-speed parameter estimation for problems with computationally expensive likelihood calculations.

We initially test \texttt{Gaussbock} on a fast approximation of the DES Y1 posterior to demonstrate its capabilities on a high-dimensional realistic example, and to investigate scaling relations and the effectiveness of the convergence criterion, both of which prove to be well-behaved. We then apply our method to the full DES Y1 posterior, making use of \texttt{Gaussbock}'s built-in MPI capabilties to run it on NERSC supercomputing facilities. The results showcase the immense speed-up that constitutes the primary strength of our method, achieving a good fit to the original DES approach of using \texttt{MultiNest}.

While achieving excellent fits in most cases across our experiments, we observe that less Gaussian posteriors of unconstrained parameters result in a slightly worse fit to the tails of the distribution and discuss the potential issues arising from less smooth posterior surfaces. In addition, we stress-test the algorithm using more complex distributions. We also demonstrate that \texttt{Gaussbock} achieves tight fits to hard-to-approximate posteriors such as double Gaussian shells, scattered multivariate Gaussians, and exponential distributions in lower dimensions. The reliance on an initial sample guess roughly covering the areas of interest, however, means that it will break down if the latter is not the case, for example if modes of a multivariate distribution are not caught in that initial sample. In addition, we verify that our method, like other parameter estimation techniques based on Gaussian mixture models, is limited by the degree to which distributions can be formalized as a weighted mixture of Gaussians, which becomes problematic if, for example, facing Gaussian shells of moderate to high dimensionality.

We implement \texttt{Gaussbock} as a pure-Python package to conduct our experiments described in this paper. In doing so, we also provide the astronomy community with a user-friendly and readily installable implementation of \texttt{Gaussbock}, bearing the same name. While our method is developed specifically with contemporary parameter estimation problems in cosmology in mind, it represents a general-purpose inference tool applicable to many problems dealing with high-dimensional parameter estimation with computationally costly posteriors. As a result, our work contributes to the wider field of estimation theory in addition to current and upcoming astronomical surveys.

\section*{Acknowledgments}

We thank members of the Dark Energy Survey collaboration for making DES data publicly available. BM acknowledges the support of a Principal's Career Development Scholarship from the University of Edinburgh. JZ acknowledges a Chancellor's Fellowship, also from the University of Edinburgh.

\bibliographystyle{aasjournal}
\bibliography{references}

\begin{thebibliography}{}
\expandafter\ifx\csname natexlab\endcsname\relax\def\natexlab#1{#1}\fi
\providecommand{\url}[1]{\href{#1}{#1}}

\bibitem[{{Abbott} {et~al.}(2016{\natexlab{a}}){Abbott}, {Abdalla},
  {Aleksi{\'c}}, {Allam}, {Amara}, {Bacon}, {Balbinot}, {Banerji}, {Bechtol},
  {Benoit-L{\'e}vy}, {Bernstein}, {Bertin}, {Blazek}, {Bonnett}, {Bridle},
  {Brooks}, {Brunner}, {Buckley-Geer}, {Burke}, {Caminha}, {Capozzi},
  {Carlsen}, {Carnero-Rosell}, {Carollo}, {Carrasco-Kind}, {Carretero},
  {Castander}, {Clerkin}, {Collett}, {Conselice}, {Crocce}, {Cunha},
  {D'Andrea}, {da Costa}, {Davis}, {Desai}, {Diehl}, {Dietrich}, {Dodelson},
  {Doel}, {Drlica-Wagner}, {Estrada}, {Etherington}, {Evrard}, {Fabbri},
  {Finley}, {Flaugher}, {Foley}, {Fosalba}, {Frieman}, {Garc{\'\i}a-Bellido},
  {Gaztanaga}, {Gerdes}, {Giannantonio}, {Goldstein}, {Gruen}, {Gruendl},
  {Guarnieri}, {Gutierrez}, {Hartley}, {Honscheid}, {Jain}, {James}, {Jeltema},
  {Jouvel}, {Kessler}, {King}, {Kirk}, {Kron}, {Kuehn}, {Kuropatkin}, {Lahav},
  {Li}, {Lima}, {Lin}, {Maia}, {Makler}, {Manera}, {Maraston}, {Marshall},
  {Martini}, {McMahon}, {Melchior}, {Merson}, {Miller}, {Miquel}, {Mohr},
  {Morice-Atkinson}, {Naidoo}, {Neilsen}, {Nichol}, {Nord}, {Ogando},
  {Ostrovski}, {Palmese}, {Papadopoulos}, {Peiris}, {Peoples}, {Percival},
  {Plazas}, {Reed}, {Refregier}, {Romer}, {Roodman}, {Ross}, {Rozo}, {Rykoff},
  {Sadeh}, {Sako}, {S{\'a}nchez}, {Sanchez}, {Santiago}, {Scarpine},
  {Schubnell}, {Sevilla-Noarbe}, {Sheldon}, {Smith}, {Smith}, {Soares-Santos},
  {Sobreira}, {Soumagnac}, {Suchyta}, {Sullivan}, {Swanson}, {Tarle}, {Thaler},
  {Thomas}, {Thomas}, {Tucker}, {Vieira}, {Vikram}, {Walker}, {Wechsler},
  {Weller}, {Wester}, {Whiteway}, {Wilcox}, {Yanny}, {Zhang}, \&
  {Zuntz}}]{Abbott2016a}
{Abbott}, T., {Abdalla}, F.~B., {Aleksi{\'c}}, J., {et~al.} 2016{\natexlab{a}},
  \mnras, 460, 1270

\bibitem[{{Abbott} {et~al.}(2016{\natexlab{b}}){Abbott}, {Abdalla}, {Allam},
  {Amara}, {Annis}, {Armstrong}, {Bacon}, {Banerji}, {Bauer}, {Baxter},
  {Becker}, {Benoit-L{\'e}vy}, {Bernstein}, {Bernstein}, {Bertin}, {Blazek},
  {Bonnett}, {Bridle}, {Brooks}, {Bruderer}, {Buckley-Geer}, {Burke}, {Busha},
  {Capozzi}, {Carnero Rosell}, {Carrasco Kind}, {Carretero}, {Castander},
  {Chang}, {Clampitt}, {Crocce}, {Cunha}, {D'Andrea}, {da Costa}, {Das},
  {DePoy}, {Desai}, {Diehl}, {Dietrich}, {Dodelson}, {Doel}, {Drlica-Wagner},
  {Efstathiou}, {Eifler}, {Erickson}, {Estrada}, {Evrard}, {Fausti Neto},
  {Fernandez}, {Finley}, {Flaugher}, {Fosalba}, {Friedrich}, {Frieman},
  {Gangkofner}, {Garcia-Bellido}, {Gaztanaga}, {Gerdes}, {Gruen}, {Gruendl},
  {Gutierrez}, {Hartley}, {Hirsch}, {Honscheid}, {Huff}, {Jain}, {James},
  {Jarvis}, {Kacprzak}, {Kent}, {Kirk}, {Krause}, {Kravtsov}, {Kuehn},
  {Kuropatkin}, {Kwan}, {Lahav}, {Leistedt}, {Li}, {Lima}, {Lin}, {MacCrann},
  {March}, {Marshall}, {Martini}, {McMahon}, {Melchior}, {Miller}, {Miquel},
  {Mohr}, {Neilsen}, {Nichol}, {Nicola}, {Nord}, {Ogando}, {Palmese}, {Peiris},
  {Plazas}, {Refregier}, {Roe}, {Romer}, {Roodman}, {Rowe}, {Rykoff}, {Sabiu},
  {Sadeh}, {Sako}, {Samuroff}, {Sanchez}, {S{\'a}nchez}, {Seo},
  {Sevilla-Noarbe}, {Sheldon}, {Smith}, {Soares-Santos}, {Sobreira}, {Suchyta},
  {Swanson}, {Tarle}, {Thaler}, {Thomas}, {Troxel}, {Vikram}, {Walker},
  {Wechsler}, {Weller}, {Zhang}, {Zuntz}, \& {Dark Energy Survey
  Collaboration}}]{Abbott2016b}
{Abbott}, T., {Abdalla}, F.~B., {Allam}, S., {et~al.} 2016{\natexlab{b}}, \prd,
  94, 022001

\bibitem[{Abbott {et~al.}(2018)Abbott, Abdalla, Alarcon,
  Aleksi\ifmmode~\acute{c}\else \'{c}\fi{}, Allam, Allen, Amara, Annis, Asorey,
  Avila, Bacon, Balbinot, Banerji, Banik, Barkhouse, Baumer, Baxter, Bechtol,
  Becker, Benoit-L\'evy, Benson, Bernstein, Bertin, Blazek, Bridle, Brooks,
  Brout, Buckley-Geer, Burke, Busha, Campos, Capozzi, Carnero~Rosell,
  Carrasco~Kind, Carretero, Castander, Cawthon, Chang, Chen, Childress, Choi,
  Conselice, Crittenden, Crocce, Cunha, D'Andrea, da~Costa, Das, Davis, Davis,
  De~Vicente, DePoy, DeRose, Desai, Diehl, Dietrich, Dodelson, Doel,
  Drlica-Wagner, Eifler, Elliott, Elsner, Elvin-Poole, Estrada, Evrard, Fang,
  Fernandez, Fert\'e, Finley, Flaugher, Fosalba, Friedrich, Frieman,
  Garc\'{\i}a-Bellido, Garcia-Fernandez, Gatti, Gaztanaga, Gerdes,
  Giannantonio, Gill, Glazebrook, Goldstein, Gruen, Gruendl, Gschwend,
  Gutierrez, Hamilton, Hartley, Hinton, Honscheid, Hoyle, Huterer, Jain, James,
  Jarvis, Jeltema, Johnson, Johnson, Kacprzak, Kent, Kim, King, Kirk, Kokron,
  Kovacs, Krause, Krawiec, Kremin, Kuehn, Kuhlmann, Kuropatkin, Lacasa, Lahav,
  Li, Liddle, Lidman, Lima, Lin, MacCrann, Maia, Makler, Manera, March,
  Marshall, Martini, McMahon, Melchior, Menanteau, Miquel, Miranda, Mudd, Muir,
  M\"oller, Neilsen, Nichol, Nord, Nugent, Ogando, Palmese, Peacock, Peiris,
  Peoples, Percival, Petravick, Plazas, Porredon, Prat, Pujol, Rau, Refregier,
  Ricker, Roe, Rollins, Romer, Roodman, Rosenfeld, Ross, Rozo, Rykoff, Sako,
  Salvador, Samuroff, S\'anchez, Sanchez, Santiago, Scarpine, Schindler,
  Scolnic, Secco, Serrano, Sevilla-Noarbe, Sheldon, Smith, Smith, Smith,
  Soares-Santos, Sobreira, Suchyta, Tarle, Thomas, Troxel, Tucker, Tucker,
  Uddin, Varga, Vielzeuf, Vikram, Vivas, Walker, Wang, Wechsler, Weller,
  Wester, Wolf, Yanny, Yuan, Zenteno, Zhang, Zhang, \& Zuntz}]{Abbott2018a}
Abbott, T. M.~C., Abdalla, F.~B., Alarcon, A., {et~al.} 2018, \prd, 98, 043526

\bibitem[{{Abbott} {et~al.}(2018){Abbott}, {Abdalla}, {Allam}, {Amara},
  {Annis}, {Asorey}, {Avila}, {Ballester}, {Banerji}, {Barkhouse}, {Baruah},
  {Baumer}, {Bechtol}, {Becker}, {Benoit-L{\'e}vy}, {Bernstein}, {Bertin},
  {Blazek}, {Bocquet}, {Brooks}, {Brout}, {Buckley-Geer}, {Burke}, {Busti},
  {Campisano}, {Cardiel-Sas}, {Carnero Rosell}, {Carrasco Kind}, {Carretero},
  {Castander}, {Cawthon}, {Chang}, {Chen}, {Conselice}, {Costa}, {Crocce},
  {Cunha}, {D{\textquoteright}Andrea}, {da Costa}, {Das}, {Daues}, {Davis},
  {Davis}, {De Vicente}, {DePoy}, {DeRose}, {Desai}, {Diehl}, {Dietrich},
  {Dodelson}, {Doel}, {Drlica-Wagner}, {Eifler}, {Elliott}, {Evrard}, {Farahi},
  {Fausti Neto}, {Fernandez}, {Finley}, {Flaugher}, {Foley}, {Fosalba},
  {Friedel}, {Frieman}, {Garc{\'\i}a-Bellido}, {Gaztanaga}, {Gerdes},
  {Giannantonio}, {Gill}, {Glazebrook}, {Goldstein}, {Gower}, {Gruen},
  {Gruendl}, {Gschwend}, {Gupta}, {Gutierrez}, {Hamilton}, {Hartley}, {Hinton},
  {Hislop}, {Hollowood}, {Honscheid}, {Hoyle}, {Huterer}, {Jain}, {James},
  {Jeltema}, {Johnson}, {Johnson}, {Kacprzak}, {Kent}, {Khullar}, {Klein},
  {Kovacs}, {Koziol}, {Krause}, {Kremin}, {Kron}, {Kuehn}, {Kuhlmann},
  {Kuropatkin}, {Lahav}, {Lasker}, {Li}, {Li}, {Liddle}, {Lima}, {Lin},
  {L{\'o}pez-Reyes}, {MacCrann}, {Maia}, {Maloney}, {Manera}, {March},
  {Marriner}, {Marshall}, {Martini}, {McClintock}, {McKay}, {McMahon},
  {Melchior}, {Menanteau}, {Miller}, {Miquel}, {Mohr}, {Morganson}, {Mould},
  {Neilsen}, {Nichol}, {Nogueira}, {Nord}, {Nugent}, {Nunes}, {Ogando}, {Old},
  {Pace}, {Palmese}, {Paz-Chinch{\'o}n}, {Peiris}, {Percival}, {Petravick},
  {Plazas}, {Poh}, {Pond}, {Porredon}, {Pujol}, {Refregier}, {Reil}, {Ricker},
  {Rollins}, {Romer}, {Roodman}, {Rooney}, {Ross}, {Rykoff}, {Sako}, {Sanchez},
  {Sanchez}, {Santiago}, {Saro}, {Scarpine}, {Scolnic}, {Serrano},
  {Sevilla-Noarbe}, {Sheldon}, {Shipp}, {Silveira}, {Smith}, {Smith}, {Smith},
  {Soares-Santos}, {Sobreira}, {Song}, {Stebbins}, {Suchyta}, {Sullivan},
  {Swanson}, {Tarle}, {Thaler}, {Thomas}, {Thomas}, {Troxel}, {Tucker},
  {Vikram}, {Vivas}, {Walker}, {Wechsler}, {Weller}, {Wester}, {Wolf}, {Wu},
  {Yanny}, {Zenteno}, {Zhang}, {Zuntz}, {DES Collaboration}, {Juneau},
  {Fitzpatrick}, {Nikutta}, {Nidever}, {Olsen}, {Scott}, \& {Data
  Lab}}]{Abbott2018b}
{Abbott}, T.~M.~C., {Abdalla}, F.~B., {Allam}, S., {et~al.} 2018, \apjs, 239,
  18

\bibitem[{{Aghanim} {et~al.}(2018){Aghanim}, {Akrami}, {Ashdown}, {Aumont},
  {Baccigalupi}, {Ballardini}, {Banday}, {Barreiro}, {Bartolo}, {Basak},
  {Battye}, {Benabed}, {Bernard}, {Bersanelli}, {Bielewicz}, {Bock}, {Bond},
  {Borrill}, {Bouchet}, {Boulanger}, {Bucher}, {Burigana}, {Butler},
  {Calabrese}, {Cardoso}, {Carron}, {Challinor}, {Chiang}, {Chluba}, {Colombo},
  {Combet}, {Contreras}, {Crill}, {Cuttaia}, {de Bernardis}, {de Zotti},
  {Delabrouille}, {Delouis}, {Di Valentino}, {Diego}, {Dor{\'e}}, {Douspis},
  {Ducout}, {Dupac}, {Dusini}, {Efstathiou}, {Elsner}, {En{\ss}lin}, {Eriksen},
  {Fantaye}, {Farhang}, {Fergusson}, {Fernandez-Cobos}, {Finelli},
  {Forastieri}, {Frailis}, {Franceschi}, {Frolov}, {Galeotta}, {Galli},
  {Ganga}, {G{\'e}nova-Santos}, {Gerbino}, {Ghosh}, {Gonz{\'a}lez-Nuevo},
  {G{\'o}rski}, {Gratton}, {Gruppuso}, {Gudmundsson}, {Hamann}, {Handley},
  {Herranz}, {Hivon}, {Huang}, {Jaffe}, {Jones}, {Karakci}, {Keih{\"a}nen},
  {Keskitalo}, {Kiiveri}, {Kim}, {Kisner}, {Knox}, {Krachmalnicoff}, {Kunz},
  {Kurki-Suonio}, {Lagache}, {Lamarre}, {Lasenby}, {Lattanzi}, {Lawrence}, {Le
  Jeune}, {Lemos}, {Lesgourgues}, {Levrier}, {Lewis}, {Liguori}, {Lilje},
  {Lilley}, {Lindholm}, {L{\'o}pez-Caniego}, {Lubin}, {Ma},
  {Mac{\'\i}as-P{\'e}rez}, {Maggio}, {Maino}, {Mandolesi}, {Mangilli},
  {Marcos-Caballero}, {Maris}, {Martin}, {Martinelli},
  {Mart{\'\i}nez-Gonz{\'a}lez}, {Matarrese}, {Mauri}, {McEwen}, {Meinhold},
  {Melchiorri}, {Mennella}, {Migliaccio}, {Millea}, {Mitra},
  {Miville-Desch{\^e}nes}, {Molinari}, {Montier}, {Morgante}, {Moss}, {Natoli},
  {N{\o}rgaard-Nielsen}, {Pagano}, {Paoletti}, {Partridge}, {Patanchon},
  {Peiris}, {Perrotta}, {Pettorino}, {Piacentini}, {Polastri}, {Polenta},
  {Puget}, {Rachen}, {Reinecke}, {Remazeilles}, {Renzi}, {Rocha}, {Rosset},
  {Roudier}, {Rubi{\~n}o-Mart{\'\i}n}, {Ruiz-Granados}, {Salvati}, {Sandri},
  {Savelainen}, {Scott}, {Shellard}, {Sirignano}, {Sirri}, {Spencer},
  {Sunyaev}, {Suur-Uski}, {Tauber}, {Tavagnacco}, {Tenti}, {Toffolatti},
  {Tomasi}, {Trombetti}, {Valenziano}, {Valiviita}, {Van Tent}, {Vibert},
  {Vielva}, {Villa}, {Vittorio}, {Wandelt}, {Wehus}, {White}, {White},
  {Zacchei}, \& {Zonca}}]{Aghanim2018}
{Aghanim}, N., {Akrami}, Y., {Ashdown}, M., {et~al.} 2018, arXiv e-prints,
  arXiv:1807.06209

\bibitem[{{Ahmadian} {et~al.}(2011){Ahmadian}, {Pillow}, \&
  {Paninski}}]{Ahmadian2011}
{Ahmadian}, Y., {Pillow}, J.~W., \& {Paninski}, L. 2011, Neural Comput., 23, 46

\bibitem[{{Aitken} \& {Akman}(2013)}]{Aitken2013}
{Aitken}, S., \& {Akman}, O.~E. 2013, BMC Syst. Biol., 7, 72

\bibitem[{{Akeret} {et~al.}(2015){Akeret}, {Refregier}, {Amara}, {Seehars}, \&
  {Hasner}}]{Akeret2015}
{Akeret}, J., {Refregier}, A., {Amara}, A., {Seehars}, S., \& {Hasner}, C.
  2015, \jcap, 2015, 043

\bibitem[{{Akeret} {et~al.}(2013){Akeret}, {Seehars}, {Amara}, {Refregier}, \&
  {Csillaghy}}]{Akeret2013}
{Akeret}, J., {Seehars}, S., {Amara}, A., {Refregier}, A., \& {Csillaghy}, A.
  2013, Astron. Comput., 2, 27

\bibitem[{{Allanach} \& {Lester}(2008)}]{Allanach2008}
{Allanach}, B., \& {Lester}, C. 2008, Computer Physics Communications, 179, 256

\bibitem[{{Allison} \& {Dunkley}(2014)}]{Allison2014}
{Allison}, R., \& {Dunkley}, J. 2014, \mnras, 437, 3918

\bibitem[{{Alsing} {et~al.}(2018){Alsing}, {Wandelt}, \& {Feeney}}]{Alsing2018}
{Alsing}, J., {Wandelt}, B., \& {Feeney}, S. 2018, \mnras, 477, 2874

\bibitem[{{Amendola} {et~al.}(2018){Amendola}, {Appleby}, {Avgoustidis},
  {Bacon}, {Baker}, {Baldi}, {Bartolo}, {Blanchard}, {Bonvin}, {Borgani},
  {Branchini}, {Burrage}, {Camera}, {Carbone}, {Casarini}, {Cropper}, {de
  Rham}, {Dietrich}, {Di Porto}, {Durrer}, {Ealet}, {Ferreira}, {Finelli},
  {Garc{\'\i}a-Bellido}, {Giannantonio}, {Guzzo}, {Heavens}, {Heisenberg},
  {Heymans}, {Hoekstra}, {Hollenstein}, {Holmes}, {Hwang}, {Jahnke},
  {Kitching}, {Koivisto}, {Kunz}, {La Vacca}, {Linder}, {March}, {Marra},
  {Martins}, {Majerotto}, {Markovic}, {Marsh}, {Marulli}, {Massey}, {Mellier},
  {Montanari}, {Mota}, {Nunes}, {Percival}, {Pettorino}, {Porciani},
  {Quercellini}, {Read}, {Rinaldi}, {Sapone}, {Sawicki}, {Scaramella},
  {Skordis}, {Simpson}, {Taylor}, {Thomas}, {Trotta}, {Verde}, {Vernizzi},
  {Vollmer}, {Wang}, {Weller}, \& {Zlosnik}}]{Amendola2018}
{Amendola}, L., {Appleby}, S., {Avgoustidis}, A., {et~al.} 2018, Living Rev.
  Relativ., 21, 2

\bibitem[{{Antoniak}(1974)}]{Antoniak1974}
{Antoniak}, C.~E. 1974, Ann. Statist., 2, 1152

\bibitem[{{Bardenet} {et~al.}(2014){Bardenet}, {Doucet}, \&
  {Holmes}}]{Bardenet2014}
{Bardenet}, R., {Doucet}, A., \& {Holmes}, C. 2014, in Proceedings of the 31st
  International Conference on Machine Learning (ICML 2014), 405--413

\bibitem[{{Bernardo} \& {Smith}(1994)}]{Bernardo1994}
{Bernardo}, J., \& {Smith}, A. F.~M. 1994, {Bayesian theory} (Hoboken: John
  Wiley \& Sons, Inc.)

\bibitem[{{Betoule} {et~al.}(2014){Betoule}, {Kessler}, {Guy}, {Mosher},
  {Hardin}, {Biswas}, {Astier}, {El-Hage}, {Konig}, {Kuhlmann}, {Marriner},
  {Pain}, {Regnault}, {Balland}, {Bassett}, {Brown}, {Campbell}, {Carlberg},
  {Cellier-Holzem}, {Cinabro}, {Conley}, {D'Andrea}, {DePoy}, {Doi}, {Ellis},
  {Fabbro}, {Filippenko}, {Foley}, {Frieman}, {Fouchez}, {Galbany}, {Goobar},
  {Gupta}, {Hill}, {Hlozek}, {Hogan}, {Hook}, {Howell}, {Jha}, {Le Guillou},
  {Leloudas}, {Lidman}, {Marshall}, {M{\"o}ller}, {Mour{\~a}o}, {Neveu},
  {Nichol}, {Olmstead}, {Palanque-Delabrouille}, {Perlmutter}, {Prieto},
  {Pritchet}, {Richmond}, {Riess}, {Ruhlmann-Kleider}, {Sako}, {Schahmaneche},
  {Schneider}, {Smith}, {Sollerman}, {Sullivan}, {Walton}, \&
  {Wheeler}}]{Betoule2014}
{Betoule}, M., {Kessler}, R., {Guy}, J., {et~al.} 2014, \aap, 568, A22

\bibitem[{{Bishop}(2006)}]{Bishop2006}
{Bishop}, C.~M. 2006, {Pattern recognition and machine learning} (Heidelberg,
  Germany: Springer-Verlag)

\bibitem[{{Blei} \& {Jordan}(2006)}]{Blei2006}
{Blei}, D.~M., \& {Jordan}, M.~I. 2006, Bayesian Anal., 1, 121

\bibitem[{{Blei} {et~al.}(2017){Blei}, {Kucukelbir}, \& {McAuliffe}}]{Blei2017}
{Blei}, D.~M., {Kucukelbir}, A., \& {McAuliffe}, J.~D. 2017, J. Am. Stat.
  Assoc., 112, 859

\bibitem[{{Bonassi} {et~al.}(2011){Bonassi}, {Lingchong}, \&
  {West}}]{Bonassi2011}
{Bonassi}, F.~V., {Lingchong}, Y., \& {West}, M. 2011, Stat. Appl. Genet. Mol.
  Biol., 10, 49

\bibitem[{{Brooks} {et~al.}(2011){Brooks}, {Gelman}, {Jones}, \&
  Xiao-Li}]{Brooks2011}
{Brooks}, S., {Gelman}, A., {Jones}, G., \& Xiao-Li, m. 2011, {Handbook of
  Markov chain Monte Carlo} (New York: Chapman \& Hall (CRC Press))

\bibitem[{{Capp{\'e}} {et~al.}(2004){Capp{\'e}}, {Guillin}, {Marin}, \&
  {Robert}}]{Cappe2004}
{Capp{\'e}}, O., {Guillin}, A., {Marin}, J.-M., \& {Robert}, C.~P. 2004, J.
  Comput. Graph. Stat., 13, 907

\bibitem[{{Chopin} \& {Robert}(2010)}]{Chopin2010}
{Chopin}, N., \& {Robert}, C. 2010, Biometrika, 97, 741

\bibitem[{{Christensen} \& {Meyer}(1998)}]{Christensen1998}
{Christensen}, N., \& {Meyer}, R. 1998, \prd, 58, 082001

\bibitem[{{Christensen} {et~al.}(2001){Christensen}, {Meyer}, {Knox}, \&
  {Luey}}]{Christensen2001}
{Christensen}, N., {Meyer}, R., {Knox}, L., \& {Luey}, B. 2001, Class. Quantum
  Grav., 18, 2677

\bibitem[{{Del Pozzo}(2012)}]{DelPozzo2012}
{Del Pozzo}, W. 2012, \prd, 86, 043011

\bibitem[{{Del Pozzo} {et~al.}(2017){Del Pozzo}, {Li}, \&
  {Messenger}}]{DelPozzo2017}
{Del Pozzo}, W., {Li}, T. G.~F., \& {Messenger}, C. 2017, \prd, 95, 043502

\bibitem[{{Duane} {et~al.}(1987){Duane}, {Kennedy}, {Pendleton}, \&
  {Roweth}}]{Duane1987}
{Duane}, S., {Kennedy}, A.~D., {Pendleton}, B.~J., \& {Roweth}, D. 1987, Phys.
  Lett. B, 195, 216

\bibitem[{{Fan} {et~al.}(2013){Fan}, {Nott}, \& {Sisson}}]{Fan2013}
{Fan}, Y., {Nott}, D.~J., \& {Sisson}, S.~A. 2013, Stat, 2, 34

\bibitem[{{Ferguson}(1973)}]{Ferguson1973}
{Ferguson}, T.~S. 1973, Ann. Statist., 1, 209

\bibitem[{{Feroz} {et~al.}(2009){Feroz}, {Hobson}, \& {Bridges}}]{Feroz2009}
{Feroz}, F., {Hobson}, M.~P., \& {Bridges}, M. 2009, \mnras, 398, 1601

\bibitem[{{Foreman-Mackey} {et~al.}(2013){Foreman-Mackey}, {Hogg}, {Lang}, \&
  {Goodman}}]{Foreman-Mackey2013}
{Foreman-Mackey}, D., {Hogg}, D.~W., {Lang}, D., \& {Goodman}, J. 2013, \pasp,
  125, 306

\bibitem[{{Gelman} {et~al.}(2013){Gelman}, {Carlin}, {Stern}, \&
  {Rubin}}]{Gelman2013}
{Gelman}, A., {Carlin}, J.~B., {Stern}, H.~S., \& {Rubin}, D.~B. 2013,
  {Bayesian data analysis} (New York: Chapman \& Hall (CRC Press))

\bibitem[{{Geman} \& {Geman}(1984)}]{Geman1984}
{Geman}, S., \& {Geman}, D. 1984, IEEE Trans. Pattern Anal. Mach. Intell., 6,
  721

\bibitem[{{Gershman} \& {Blei}(2012)}]{Gershman2012}
{Gershman}, S.~J., \& {Blei}, D.~M. 2012, J. Math. Psychol., 56, 1

\bibitem[{{Goodman} \& {Weare}(2010)}]{Goodman2010}
{Goodman}, J., \& {Weare}, J. 2010, Commun. Appl. Math. Comput. Sci., 5, 65

\bibitem[{{Hajian}(2007)}]{Hajian2007}
{Hajian}, A. 2007, \prd, 75, 083525

\bibitem[{{Handley} {et~al.}(2015){Handley}, {Hobson}, \&
  {Lasenby}}]{Handley2015}
{Handley}, W.~J., {Hobson}, M.~P., \& {Lasenby}, A.~N. 2015, \mnras, 453, 4384

\bibitem[{{Hastings}(1970)}]{Hastings1970}
{Hastings}, W.~K. 1970, Biometrika, 57, 97

\bibitem[{{He} {et~al.}(2018){He}, {Cook}, {Deslippe}, {Friesen}, {Gerber},
  {Hartman-Baker}, {Koniges}, {Kurth}, {Leak}, {Yang}, {Zhao}, \&
  {Hauschildt}}]{He2018}
{He}, Y., {Cook}, B., {Deslippe}, J., {et~al.} 2018, Concur. and Computat.:
  Practice and Experience, 30, e4291

\bibitem[{{Higson} {et~al.}(2018){Higson}, {Handley}, {Hobson}, \&
  {Lasenby}}]{Higson2018}
{Higson}, E., {Handley}, W., {Hobson}, M., \& {Lasenby}, A. 2018, Bayesian
  Anal., 13, 873

\bibitem[{{Hjort} {et~al.}(2010){Hjort}, {Holmes}, {Mueller}, \&
  {Walker}}]{Hjort2010}
{Hjort}, N.~L., {Holmes}, C., {Mueller}, P., \& {Walker}, S.~G. 2010, {Bayesian
  nonparametrics: Principles and practice} (Cambridge, UK: Cambridge University
  Press)

\bibitem[{Hobson \& Feroz(2008)}]{Hobson2008}
Hobson, M.~P., \& Feroz, F. 2008, \mnras, 384, 449

\bibitem[{{Hobson} {et~al.}(2009){Hobson}, {Jaffe}, {Liddle}, {Mukherjee}, \&
  {Parkinson}}]{Hobson2009}
{Hobson}, M.~P., {Jaffe}, A.~H., {Liddle}, A.~R., {Mukherjee}, P., \&
  {Parkinson}, D. 2009, {Bayesian methods in cosmology} (Cambridge, UK:
  Cambridge University Press)

\bibitem[{{Hoffmann} \& {Gelman}(2014)}]{Hoffmann2014}
{Hoffmann}, M.~D., \& {Gelman}, A. 2014, J. Mach. Learn. Res., 15, 1593

\bibitem[{{Howlett} {et~al.}(2012){Howlett}, {Lewis}, {Hall}, \&
  {Challinor}}]{Howlett2012}
{Howlett}, C., {Lewis}, A., {Hall}, A., \& {Challinor}, A. 2012, \jcap, 2012,
  027

\bibitem[{{Ionides}(2008)}]{Ionides2008}
{Ionides}, E.~L. 2008, J. Comput. Graph. Stat., 17, 295

\bibitem[{{Ishida} {et~al.}(2015){Ishida}, {Vitenti}, {Penna-Lima}, {Cisewski},
  {de Souza}, {Trindade}, {Cameron}, {Busti}, \& {COIN
  Collaboration}}]{Ishida2015}
{Ishida}, E. E.~O., {Vitenti}, S. D.~P., {Penna-Lima}, M., {et~al.} 2015,
  Astron. Comput., 13, 1

\bibitem[{{Jordan} {et~al.}(1999){Jordan}, {Ghahramani}, {Jaakkola}, \&
  {Saul}}]{Jordan1999}
{Jordan}, M.~I., {Ghahramani}, Z., {Jaakkola}, T.~S., \& {Saul}, L.~K. 1999,
  Mach. Learn., 37, 183

\bibitem[{{Kacprzak} {et~al.}(2016){Kacprzak}, {Kirk}, {Friedrich}, {Amara},
  {Refregier}, {Marian}, {Dietrich}, {Suchyta}, {Aleksi{\'c}}, {Bacon},
  {Becker}, {Bonnett}, {Bridle}, {Chang}, {Eifler}, {Hartley}, {Huff},
  {Krause}, {MacCrann}, {Melchior}, {Nicola}, {Samuroff}, {Sheldon}, {Troxel},
  {Weller}, {Zuntz}, {Abbott}, {Abdalla}, {Armstrong}, {Benoit-L{\'e}vy},
  {Bernstein}, {Bernstein}, {Bertin}, {Brooks}, {Burke}, {Carnero Rosell},
  {Carrasco Kind}, {Carretero}, {Castander}, {Crocce}, {D'Andrea}, {da Costa},
  {Desai}, {Diehl}, {Evrard}, {Neto}, {Flaugher}, {Fosalba}, {Frieman},
  {Gerdes}, {Goldstein}, {Gruen}, {Gruendl}, {Gutierrez}, {Honscheid}, {Jain},
  {James}, {Jarvis}, {Kuehn}, {Kuropatkin}, {Lahav}, {Lima}, {March},
  {Marshall}, {Martini}, {Miller}, {Miquel}, {Mohr}, {Nichol}, {Nord},
  {Plazas}, {Romer}, {Roodman}, {Rykoff}, {Sanchez}, {Scarpine}, {Schubnell},
  {Sevilla-Noarbe}, {Smith}, {Soares-Santos}, {Sobreira}, {Swanson}, {Tarle},
  {Thomas}, {Vikram}, {Walker}, {Zhang}, \& {DES Collaboration}}]{Kacprzak2016}
{Kacprzak}, T., {Kirk}, D., {Friedrich}, O., {et~al.} 2016, \mnras, 463, 3653

\bibitem[{{Kahn} \& {Marshall}(1953)}]{Kahn1953}
{Kahn}, H., \& {Marshall}, A.~W. 1953, Oper. Res., 1, 263

\bibitem[{{Keeton}(2011)}]{Keeton2011}
{Keeton}, C.~R. 2011, \mnras, 414, 1418

\bibitem[{{Kilbinger} {et~al.}(2009){Kilbinger}, {Benabed}, {Guy}, {Astier},
  {Tereno}, {Fu}, {Wraith}, {Coupon}, {Mellier}, {Balland}, {Bouchet},
  {Hamana}, {Hardin}, {McCracken}, {Pain}, {Regnault}, {Schultheis}, \&
  {Yahagi}}]{Kilbinger2009}
{Kilbinger}, M., {Benabed}, K., {Guy}, J., {et~al.} 2009, \aap, 497, 677

\bibitem[{{Kilbinger} {et~al.}(2010){Kilbinger}, {Wraith}, {Robert}, {Benabed},
  {Capp{\'e}}, {Cardoso}, {Fort}, {Prunet}, \& {Bouchet}}]{Kilbinger2010}
{Kilbinger}, M., {Wraith}, D., {Robert}, C.~P., {et~al.} 2010, \mnras, 405,
  2381

\bibitem[{{Knox} {et~al.}(2001){Knox}, {Christensen}, \& {Skordis}}]{Knox2001}
{Knox}, L., {Christensen}, N., \& {Skordis}, C. 2001, \apj, 563, L95

\bibitem[{{Krause} \& {Eifler}(2017)}]{Krause2017}
{Krause}, E., \& {Eifler}, T. 2017, \mnras, 470, 2100

\bibitem[{{Krause} {et~al.}(2017){Krause}, {Eifler}, {Zuntz}, {Friedrich},
  {Troxel}, {Dodelson}, {Blazek}, {Secco}, {MacCrann}, {Baxter}, {Chang},
  {Chen}, {Crocce}, {DeRose}, {Ferte}, {Kokron}, {Lacasa}, {Miranda}, {Omori},
  {Porredon}, {Rosenfeld}, {Samuroff}, {Wang}, {Wechsler}, {Abbott}, {Abdalla},
  {Allam}, {Annis}, {Bechtol}, {Benoit-Levy}, {Bernstein}, {Brooks}, {Burke},
  {Capozzi}, {Carrasco Kind}, {Carretero}, {D'Andrea}, {da Costa}, {Davis},
  {DePoy}, {Desai}, {Diehl}, {Dietrich}, {Evrard}, {Flaugher}, {Fosalba},
  {Frieman}, {Garcia-Bellido}, {Gaztanaga}, {Giannantonio}, {Gruen}, {Gruendl},
  {Gschwend}, {Gutierrez}, {Honscheid}, {James}, {Jeltema}, {Kuehn},
  {Kuhlmann}, {Lahav}, {Lima}, {Maia}, {March}, {Marshall}, {Martini},
  {Menanteau}, {Miquel}, {Nichol}, {Plazas}, {Romer}, {Rykoff}, {Sanchez},
  {Scarpine}, {Schindler}, {Schubnell}, {Sevilla-Noarbe}, {Smith},
  {Soares-Santos}, {Sobreira}, {Suchyta}, {Swanson}, {Tarle}, {Tucker},
  {Vikram}, {Walker}, \& {Weller}}]{Krause2018}
{Krause}, E., {Eifler}, T.~F., {Zuntz}, J., {et~al.} 2017, arXiv e-prints,
  arXiv:1706.09359

\bibitem[{{Lewis} \& {Bridle}(2002)}]{Lewis2002}
{Lewis}, A., \& {Bridle}, S. 2002, \prd, 66, 103511

\bibitem[{{Liddle} {et~al.}(2006){Liddle}, {Parkinson}, \&
  {Mukherjee}}]{Liddle2006}
{Liddle}, A., {Parkinson}, D., \& {Mukherjee}, P. 2006, Astron. Geophys., 47,
  4.30

\bibitem[{{MacKay}(2003)}]{MacKay2003}
{MacKay}, D. J.~C. 2003, {Information theory, inference and learning
  algorithms} (Cambridge, UK: Cambridge University Press)

\bibitem[{{Metropolis} {et~al.}(1953){Metropolis}, {Rosenbluth}, {Rosenbluth},
  {Teller}, \& {Teller}}]{Metropolis1953}
{Metropolis}, N., {Rosenbluth}, A.~W., {Rosenbluth}, M.~N., {Teller}, A.~H., \&
  {Teller}, E. 1953, \jcp, 21, 1087

\bibitem[{{Mukherjee} {et~al.}(2006){Mukherjee}, {Parkinson}, \&
  {Liddle}}]{Mukherjee2006}
{Mukherjee}, P., {Parkinson}, D., \& {Liddle}, A.~R. 2006, \apj, 638, L51

\bibitem[{{Murphy}(2012)}]{Murphy2012}
{Murphy}, K.~P. 2012, {Machine learning: A probabilistic perspective}
  (Cambridge, USA: The MIT Press)

\bibitem[{{Neiswanger} {et~al.}(2014){Neiswanger}, {Wang}, \&
  {Xing}}]{Neiswanger2014}
{Neiswanger}, W., {Wang}, C., \& {Xing}, E.~P. 2014, in Proceedings of the 30th
  Conference on Uncertainty in Artificial Intelligence (UAI'14), 623--632

\bibitem[{{O'Brien} {et~al.}(2016){O'Brien}, {Kashinath}, {Cavanaugh},
  {Collins}, \& {O'Brien}}]{OBrien2016}
{O'Brien}, T.~A., {Kashinath}, K., {Cavanaugh}, N.~R., {Collins}, W.~D., \&
  {O'Brien}, J.~P. 2016, Comput. Stat. Data Anal., 101, 148

\bibitem[{{Papamakarios} \& {Murray}(2016)}]{Papamakarios2016}
{Papamakarios}, G., \& {Murray}, I. 2016, in Advances in Neural Information
  Processing Systems 29, ed. D.~D. Lee, M.~Sugiyama, U.~V. Luxburg, I.~Guyon,
  \& R.~Garnett (Red Hook, USA: Curran Associates), 1028--1036

\bibitem[{{Peterson} \& {Anderson}(1987)}]{Peterson1987}
{Peterson}, C., \& {Anderson}, J.~R. 1987, Complex Syst., 1, 995

\bibitem[{{Peterson} \& {Hartman}(1989)}]{Peterson1989}
{Peterson}, C., \& {Hartman}, E. 1989, Neural Netw., 2, 475

\bibitem[{{Price-Whelan} \& {Foreman-Mackey}(2017)}]{Price-Whelan2017}
{Price-Whelan}, A.~M., \& {Foreman-Mackey}, D. 2017, J. Open Source Softw., 2,
  357

\bibitem[{Robert \& Casella(2011)}]{Robert2011}
Robert, C., \& Casella, G. 2011, Statistical Science, 26, 102

\bibitem[{{Robert} \& {Casella}(2004)}]{Robert2004}
{Robert}, C.~P., \& {Casella}, G. 2004, {Monte Carlo statistical methods}
  (Heidelberg, Germany: Springer-Verlag)

\bibitem[{{Robert} {et~al.}(2018){Robert}, {Elvira}, {Tawn}, \&
  {Wu}}]{Robert2018}
{Robert}, C.~P., {Elvira}, V., {Tawn}, N., \& {Wu}, C. 2018, WIREs Comput.
  Stat., 10, e1435

\bibitem[{{Saha} \& {Williams}(1994)}]{Saha1994}
{Saha}, P., \& {Williams}, T.~B. 1994, \aj, 107, 1295

\bibitem[{{Sethuraman}(1994)}]{Sethuraman1994}
{Sethuraman}, J. 1994, Stat. Sin., 4, 639

\bibitem[{{Skilling}(2006)}]{Skilling2006}
{Skilling}, J. 2006, Bayesian Anal., 1, 833

\bibitem[{{Smith} {et~al.}(2003){Smith}, {Peacock}, {Jenkins}, {White},
  {Frenk}, {Pearce}, {Thomas}, {Efstathiou}, \& {Couchman}}]{Smith2003}
{Smith}, R.~E., {Peacock}, J.~A., {Jenkins}, A., {et~al.} 2003, \mnras, 341,
  1311

\bibitem[{{Stumpf} {et~al.}(2014){Stumpf}, {Kirk}, \& {Johnson}}]{Stumpf2014}
{Stumpf}, M. P.~H., {Kirk}, P., \& {Johnson}, R. 2014, Bioinformatics, 31, 604

\bibitem[{{Takahashi} {et~al.}(2012){Takahashi}, {Sato}, {Nishimichi},
  {Taruya}, \& {Oguri}}]{Takahashi2012}
{Takahashi}, R., {Sato}, M., {Nishimichi}, T., {Taruya}, A., \& {Oguri}, M.
  2012, \apj, 761, 152

\bibitem[{{Thorndike}(1953)}]{Thorndike1953}
{Thorndike}, R.~L. 1953, Psychometrika, 18, 267

\bibitem[{{Torrie} \& {Valleau}(1977)}]{Torrie1977}
{Torrie}, G.~M., \& {Valleau}, J.~P. 1977, J. Comput. Phys., 23, 187

\bibitem[{{Trotta}(2008)}]{Trotta2008}
{Trotta}, R. 2008, Contemp. Phys., 49, 71

\bibitem[{{Trotta} {et~al.}(2011){Trotta}, {J{\'o}hannesson}, {Moskalenko},
  {Porter}, {Ruiz de Austri}, \& {Strong}}]{Trotta2011}
{Trotta}, R., {J{\'o}hannesson}, G., {Moskalenko}, I.~V., {et~al.} 2011, \apj,
  729, 106

\bibitem[{{Verde} {et~al.}(2013){Verde}, {Feeney}, {Mortlock}, \&
  {Peiris}}]{Verde2013}
{Verde}, L., {Feeney}, S.~M., {Mortlock}, D.~J., \& {Peiris}, H.~V. 2013,
  \jcap, 2013, 013

\bibitem[{{Wang} {et~al.}(2018){Wang}, {Wang}, \& {Xia}}]{Wang2018}
{Wang}, S., {Wang}, Y.-F., \& {Xia}, D.-M. 2018, Chin. Phys. C, 42, 065103

\bibitem[{{Wilkinson}(2005)}]{Wilkinson2005}
{Wilkinson}, D.~J. 2005, in Handbook of parallel computing and statistics (New
  York: Chapman \& Hall (CRC Press)), 477--513

\bibitem[{{Wolpert} \& {Macready}(1997)}]{Wolpert1997}
{Wolpert}, D.~H., \& {Macready}, W.~G. 1997, IEEE Trans. Evol. Comput., 1, 67

\bibitem[{{Wraith} {et~al.}(2009){Wraith}, {Kilbinger}, {Benabed}, {Capp{\'e}},
  {Cardoso}, {Fort}, {Prunet}, \& {Robert}}]{Wraith2009}
{Wraith}, D., {Kilbinger}, M., {Benabed}, K., {et~al.} 2009, \prd, 80, 023507

\bibitem[{{Zuntz} {et~al.}(2015){Zuntz}, {Paterno}, {Jennings}, {Rudd},
  {Manzotti}, {Dodelson}, {Bridle}, {Sehrish}, \& {Kowalkowski}}]{Zuntz2015}
{Zuntz}, J., {Paterno}, M., {Jennings}, E., {et~al.} 2015, Astron. Comput., 12,
  45

\end{thebibliography}

\end{document}